\documentclass[10pt, conference, compsocconf]{IEEEtran}

\usepackage[pdftex]{graphicx}
\usepackage{amsmath}
\usepackage{array}
\usepackage{listings, pdfcomment}
\usepackage{multicol}
\usepackage{multirow}
\usepackage{subfig}
\usepackage{caption}
\usepackage[linesnumbered, ruled]{algorithm2e}
\usepackage{hyperref}
\usepackage{url}

\begin{document}

\title{Are Clouds Ready to Accelerate Ad hoc Financial Simulations?}

\author{\IEEEauthorblockN{Blesson Varghese and Adam Barker}
\IEEEauthorblockA{School of Computer Science, University of St Andrews, UK\\
Email: \{varghese, adam.barker\}@st-andrews.ac.uk
}}

\maketitle

\begin{abstract}
Applications employed in the financial services industry to capture and estimate a variety of risk metrics are underpinned by stochastic simulations which are data, memory and computationally intensive. Many of these simulations are routinely performed on production-based computing systems. Ad hoc simulations in addition to routine simulations are required to obtain up-to-date views of risk metrics. Such simulations are currently not performed as they cannot be accommodated on production clusters, which are typically over committed resources. Scalable, on-demand and pay-as-you go Virtual Machines (VMs) offered by the cloud are a potential platform to satisfy the data, memory and computational constraints of the simulation. However, ``Are clouds ready to accelerate ad hoc financial simulations?'' 

The research reported in this paper aims to experimentally verify this question by developing and deploying an important financial simulation, referred to as `Aggregate Risk Analysis' on the cloud. Parallel techniques to improve efficiency and performance of the simulations are explored. Challenges such as accommodating large input data on limited memory VMs and rapidly processing data for real-time use are surmounted. The key result of this investigation is that Aggregate Risk Analysis can be accommodated on cloud VMs. Acceleration of up to 24x using multiple hardware accelerators over the implementation on a single accelerator, 6x over a multiple core implementation and approximately 60x over a baseline implementation was achieved on the cloud. However, computational time is wasted for every dollar spent on the cloud due to poor acceleration over multiple virtual cores. Interestingly, private VMs can offer better performance than public VMs on comparable underlying hardware.  

\end{abstract}


\IEEEpeerreviewmaketitle

\section{Introduction}
Financial applications in the risk industry are underpinned by large-scale stochastic simulations which are data, memory and computationally intensive \cite{simulation-1}. These simulations are run on a weekly, monthly or quarterly basis on production systems such as commodity clusters to generate risk metrics including Probable Maximum Loss (PML) \cite{pml1} and Tail Value-at-Risk (TVaR) \cite{tvar2} due to catastrophic events such as earthquakes, hurricanes and floods. The results obtained from these simulations are then interpreted by actuaries for key decision making and planning a financial year.  

The simulations that run on a routine basis are sufficient if the risk metrics do not have to be updated before their next run. Consider a simulation that takes into account the fluctuation of one parameter, for example, currency, on a weekly basis. The simulation is performed on a weekly basis to update the risk metrics. However, this is not the case in real-time scenarios where risk metrics will need to be obtained on an ad hoc basis before the next routine run. For example, consider real-time online pricing when an underwriter needs to close a deal with a client over the telephone. 

Production systems are not the right type of infrastructure for simulating on an ad hoc basis since they are firstly optimised to run routine simulations and accommodate data of known sizes, and secondly, over committed and at best fully utilised resources with no scope to satisfy the data, memory and computational requirements of ad hoc simulations. Consequentially, even if ad hoc simulations are performed on production clusters they tend to be slow. One solution to this problem would be to use dedicated systems for ad hoc simulations. However, this is not always possible since there is an additional investment on top of the maintenance costs of production clusters. An alternative solution to reduce the cost of investment is by using hardware accelerators as coprocessors in heterogeneous clusters \cite{hetcluster-1, hetcluster-2}. Though computation can be accelerated to suit ad hoc simulations the memory and data requirements cannot be always satisfied. Hardware accelerators have limited memory and thereby cannot handle large data in memory.  

Cloud computing infrastructure has the potential to address the above challenges. Maintenance costs can be eliminated and resources can be scaled on-demand, which can satisfy the requirements of ad hoc risk simulations \cite{cloud-1, cloud-3}. In this paper, the research question, ``Are clouds ready to accelerate ad hoc financial simulations?'' is explored. One application, namely Aggregate Risk Analysis, widely employed in the risk industry is developed and deployed on the cloud. Parallel techniques to accelerate the analysis and techniques to efficiently accommodate data and handle memory on cloud VMs are investigated. The experimental studies on the cloud indicate that the application achieves up to a 60x acceleration on VMs with hardware accelerators but with poor acceleration due to wasted computational time per dollar spent. Nevertheless the cloud can accommodate financial simulations. 

The remainder of this paper is organised as follows. Section \ref{relatedwork} considers research related to risk applications. Section \ref{aggregateriskanalysis} presents Aggregate Risk Analysis. Section \ref{experimentalstudies} presents an experimental study of sequential and parallel implementations on cloud VMs. This paper concludes in Section \ref{conclusion} by considering future work. 

\section{Related Work}
\label{relatedwork}
The domain of computational finance and risk addresses problems related to achieving fast computations, surmounting challenges of data management and efficiently handling memory of computing systems. Therefore, this domain is dependent on the advances in high-performance computing. Research of financial applications on production-based computing systems have progressed from small scale clusters \cite{smallcluster1, smallcluster2} to large supercomputers \cite{supercomputer1, supercomputer2}, and the typical problem addressed is achieving fast computations. These applications are hosted either on in-house clusters or on supercomputing infrastructures to which the owners of the application have access. 

A number of financial applications are being migrated from small clusters to be hosted on multiple core processors and many core coprocessor which are available at a low budget \cite{budgetplatform1}. For example, research related to financial applications exploiting the Cell BE processor  \cite{cellbe-1} \cite{cellbe-2}, FPGAs \cite{fpga1} \cite{fpga2} and GPUs \cite{gpu1, gpu2, gpu3, gpu5}. In all the above research, the need for speeding up financial applications are presented and is achieved. However, ad hoc analytics in financial risk is important which is now possible with the availability of scalable on-demand VMs provided by cloud computing and the development of big data techniques. Given that the cloud is a potential high-performance computing platform to address big data problems it is now ripe to explore risk applications in the cloud context \cite{onlinerisk}. There is limited research exploring the feasibility of accelerating and accommodating financial simulations for ad hoc analysis on the clouds. The research reported in this paper is motivated towards this end. 

\section{Aggregate Risk Analysis on the Cloud}
\label{aggregateriskanalysis}
Financial applications are underpinned by large-scale simulations which are both data, memory and computationally intensive. One such simulation is a Monte-Carlo like simulation performed on a portfolio of risk held by a reinsurer, referred to as Aggregate Risk Analysis \cite{s1, s2, s4}. This simulation provides actuaries and decision makers with millions of alternate views of catastrophic events, such as earthquakes, that can occur and the order in which they can occur in a year for portfolio risk management and real-time pricing. Millions of trials are simulated, with each trial comprising a set of possible future earthquake events and the probable loss for each trial is estimated. 

Although Aggregate Risk Analysis is an embarrassingly parallel problem, there are significant challenges in achieving efficient parallelism. One major challenge is sharing large input data between the processing cores constrained by limited memory bandwidth. Further, the challenge of accommodating input data in limited memory hardware is constrained by the complex memory architecture of accelerating hardware such as GPUs. 

Large and small sized data along with metadata are required for performing Aggregate Risk Analysis. The large data required is the Year Event Table, denoted as $YET$, which contains the occurrence of earthquake events for a year. The YET is obtained from a catalogue of possible future earthquakes that is generated using earthquake models. The frequency and the physical characteristic of the potential earthquake, and the damage the earthquake will cause are estimated by the hazard and vulnerability modules respectively of the earthquake model. The YET provides a million distinct views of potential earthquakes that can occur in a year.

Each record in a YET is called a Trial, denoted as $T_i$, which represents a possible sequence of event occurrences for any given year. The sequence of events is defined by a set of tuples containing the ID of an event and the time-stamp of its occurrence in a trial, $T_i = \{(E_{i, 1}, t_{i, 1}), \dots, (E_{i, k}, t_{i, k})\}$. The set is ordered by ascending time-stamp values. A typical YET may comprise one million trials, and each trial may have one thousand event time-stamp pairs. The YET is represented as
\begin{equation*}
\begin{array}{{l c l}}
YET	&	=	& \{ T_i = \{(E_{i, 1}, t_{i, 1}), \dots, (E_{i, k}, t_{i, k})\} \},
\end{array}
\end{equation*}
\begin{center}
where $i = 1, 2, \dots$ and $k = 1, 2, \dots, 800, \dots, 1500$.
\end{center}

The small data required for Aggregate Risk Analysis is the Event Loss Tables, denoted as $ELT$, which represents the collection of specific events and their corresponding losses. Each record in an ELT is denoted using Event-Loss pairs $EL_{i} = \{E_{i}, l_{i}\}$ and a set of financial terms associated with the ELT $\mathcal{FT}_{1} =(\mathcal{FT}_{1_{1}}, \mathcal{FT}_{1_{2}}, \dots)$. A typical Aggregate Risk Analysis may comprise ten thousand ELTs, each containing tens of thousands of event losses with exceptions even up to a few million event losses. The ELTs is represented as 
\begin{equation*}
ELT=\left\{
	\begin{array}{l c l}
	EL_{i}			&	=	&	\{E_{i}, l_{i}\},\\
	\mathcal{FT}_{1} 	&	=	&	(\mathcal{FT}_{1_{1}}, \mathcal{FT}_{1_{2}}, \dots)
	\end{array}\right\}
\end{equation*}
\begin{center}
where $i = 1, 2, \dots, 10,000, \dots, 30,000$.
\end{center}

The metadata is defined as a Portfolio, denoted as $PF$, which contains a group of Programs, denoted as $P$ represented as $PF = \{P_{1}, P_{2}, \cdots, P_{n}\}$, with $n = 1, 2, \dots, 10$. Each Program in turn covers a set of Layers, denoted as $L$, that cover a collection of ELTs under a set of financial terms of the Layer. A single layer $L_i$ is composed of the set of ELTs $\mathcal{E} = \{ELT_1, ELT_2, \dots, ELT_j\}$, and two set of Layer Terms, denoted as $\mathcal{FT}_{2} = (\mathcal{FT}_{2_{1}}, \mathcal{FT}_{2_{2}})$ and $\mathcal{FT}_{3} = (\mathcal{FT}_{3_{1}}, \mathcal{FT}_{3_{2}})$.

A typical Layer covers approximately three to thirty individual ELTs. The Layer can be represented as 
\begin{equation*}
L=\left\{
	\begin{array}{l c l}
	\mathcal{E}		& = & \{ELT_1, ELT_2, \dots, ELT_j\}, \\
	\mathcal{FT}_{2}	& = & (\mathcal{FT}_{2_{1}}, \mathcal{FT}_{2_{2}}),\\
	\mathcal{FT}_{3}	& = & (\mathcal{FT}_{3_{1}}, \mathcal{FT}_{3_{2}})
	\end{array}\right\}
\end{equation*}
\begin{center}
where $j = 1, 2, 3, \dots, 30$.
\end{center}

%\subsection{Parallel Methods for Risk Analysis}
The algorithm (line no. 1-17 shown in Algorithm \ref{algorithm1}) for aggregate analysis has two stages. In the first stage, data is loaded into local memory what is referred to as the preprocessing stage in this paper. In this stage $YET$, $ELT$ and $PF$, are loaded into memory.
\begin{algorithm}
\caption{Aggregate Risk Analysis}
\label{algorithm1}
\SetAlgoLined
\DontPrintSemicolon

\SetKwInOut{Input}{Input}
\SetKwInOut{Output}{Output}

\BlankLine

\Input{$YET$, $ELT$, $PF$}
\Output{$YLT$}

\BlankLine

\For{each Program, $P$, in $PF$}{
	\For{each Layer, $L$, in $P$}{
		\For{each Trial, $T$, in $YET$}{
			\For{each Event, $E$, in $T$}{
				\For{each $ELT$ covered by $L$}{
					Lookup $E$ in the $ELT$ and find corresponding loss, $l_{E}$\;
					Apply Financial Terms $\mathcal{FT}_{1}$ to $l_{E}$\;
					$l_{T} \leftarrow$ $l_{T}$ + $l_{E}$\;
				}
				Apply Financial Terms $\mathcal{FT}_{2}$ to $l_{T}$\;
				Apply Financial Terms $\mathcal{FT}_{3}$ to $l_{T}$\;
			}
		}
	}
}			
Populate $YLT$ using $l_{T}$\;
\BlankLine
\end{algorithm}

In the second stage, the four step simulation executed for each Layer and for each trial in the YET is performed as shown below and the resulting Year Loss Table ($YLT$) is produced.

In the first step shown in line no. 6 in which each event of a trial its corresponding event loss in the set of ELTs associated with the Layer is determined. In the second step shown in line nos. 7-9, secondary uncertainty is applied to each loss value of the Event-Loss pair extracted from an ELT. A set of contractual financial terms are then applied to the benefit of the layer. For this the losses for a specific event's net of financial terms $\mathcal{I}$ are accumulated across all ELTs into a single event loss shown in line no. 9. In the third step in line no. 11 the event loss for each event occurrence in the trial, combined across all ELTs associated with the layer, is subject to occurrence terms. In the fourth step in line no. 12 aggregate terms are applied. 

The financial terms $\mathcal{FT}_{2}$ and $\mathcal{FT}_{3}$ applied on the loss values combined across all ELTs associated with the layer are Occurrence and Aggregate terms. Two occurrence terms, namely (i) Occurrence Retention, denoted as $\mathcal{FT}_{2_{1}}$, which is the retention or deductible of the insured for an individual occurrence loss, and (ii) Occurrence Limit, denoted as $\mathcal{FT}_{2_{1}}$, which is the limit or coverage the insurer will pay for occurrence losses in excess of the retention are applied. Occurrence terms are applicable to individual event occurrences independent of any other occurrences in the trial. The occurrence terms capture specific contractual properties of 'eXcess of Loss' \cite{excessofloss-1} treaties as they apply to individual event occurrences only. The event losses net of occurrence terms are then accumulated into a single aggregate loss for the given trial. The occurrence terms are applied as $l_{T} = min ( max ( l_{T} - \mathcal{FT}_{2_{1}} ), \mathcal{FT}_{2_{2}})$.

Two aggregate terms, namely (i) Aggregate Retention, denoted as $\mathcal{FT}_{3_{1}}$, which is the retention or deductible of the insured for an annual cumulative loss, and (ii) Aggregate Limit, denoted as $\mathcal{FT}_{3_{2}}$, which is the limit or coverage the insurer will pay for annual cumulative losses in excess of the aggregate retention are applied. Aggregate terms are applied to the trial's aggregate loss for a layer. Unlike occurrence terms, aggregate terms are applied to the cumulative sum of occurrence losses within a trial and thus the result depends on the sequence of prior events in the trial. This behaviour captures contractual properties as they apply to multiple event occurrences. The aggregate loss net of the aggregate terms is referred to as the trial loss or the year loss. The aggregate terms are applied as $l_{T} = min ( max ( l_{T} - \mathcal{FT}_{3_{1}} ), \mathcal{FT}_{3_{2}})$.

The analysis generates loss values associated with each trial of the YET which populates the Year Loss Table (YLT). Important risk metrics such as the Probable Maximum Loss (PML) and the Tail Value-at-Risk (TVaR) which are used for both internal risk management and reporting to financial regulators and rating agencies can be derived from the output of the analysis. Furthermore, these metrics flow into a final stage of the risk analytics pipeline, namely Enterprise Risk Management, where liability, asset, and other forms of risks are combined and correlated to generate an enterprise wide view of risk. Additional functions can be used to generate reports that will aid actuaries and decision makers, for example, reports presenting Return Period Losses (RPL). The simulations can also be extended for providing vital information required for disaster recovery and management. 

\section{Experimental Studies}
\label{experimentalstudies}
In this section, the cloud computing platform investigated for implementing sequential and parallel Aggregate Risk Analysis is presented. The data structures chosen for representing the input and intermediate data to address memory bottlenecks and optimisations to improve the performance of the analysis are considered. An empirical study of the analysis both sequentially and in parallel on (single and multiple core) CPU and (single and multiple) GPU VMs on the cloud are presented. 

\subsection{Platform}
Aggregate risk analysis was performed on public VMs available from the Amazon Elastic Compute Cloud (EC2)\footnote{\url{http://aws.amazon.com/ec2/}} and on private VMs. Table \ref{table1} shows the underlying hardware and specifications of the VMs used for this research.

The VMs offered by Amazon are referred to as instances. Five types of instances are offered by Amazon, namely the (i) general purpose, (ii) memory optimised, (iii) cluster compute, (iv) storage optimised, and (v) GPU instances are used for the analysis. Only instances with at least 15 GB of RAM are used (this is a requirement for the analysis).

The general purpose instances employed are the \texttt{m1} and \texttt{m3} instances, namely \texttt{m1.xlarge}, \texttt{m3.xlarge} and \texttt{m3.2xlarge}. The \texttt{m1} instance is a VM of Intel(R) Xeon(R) CPU E5-2650 and \texttt{m3} instances are VMs abstracted over Intel(R) Xeon(R) CPU E5-2670. Each virtual CPU (vCPU) of the \texttt{m3} instances is a hardware hyperthread on the underlying processor. 

The memory optimised instances employed are the \texttt{m2} and \texttt{cr1} instances, namely \texttt{m2.xlarge}, \texttt{m2.2xlarge}, \texttt{m2.4xlarge} and \texttt{cr1.8xlarge}. The \texttt{m2} instances are VMs of Intel(R) Xeon(R) CPU E5-2665 and the \texttt{cr1} instance abstracts Intel(R) Xeon(R) CPU E5-2670. Each virtual CPU (vCPU) of the \texttt{cr1} instance is a hardware hyperthread on the underlying processor. 

The compute optimised instance employed are the \texttt{cc1} and \texttt{cc2} instances, namely \texttt{cc1.4xlarge} and \texttt{cc2.8xlarge}. Both \texttt{cc1} and \texttt{cc2} instances are abstraction of Intel(R) Xeon(R) CPU X5570. Each virtual CPU (vCPU) of the \texttt{cc2} instance is a hardware hyperthread on the underlying processor.  

The storage optimised instance employed are the \texttt{hi1} and \texttt{hs1} instances, namely \texttt{hi1.4xlarge} and \texttt{hs1.8xlarge}. The \texttt{hi1} instance abstracts Intel(R) Xeon(R) CPU E5620 and \texttt{hs4} is a VM over the Intel(R) Xeon(R) CPU E5-2650.

The GPU instance employed is \texttt{cg1.4xlarge} backed by two Intel(R) Xeon(R) CPU X5570 and two NVIDIA Tesla M2050 GPUs. Each GPU consists 448 processor cores and 3 GB of total global memory yielding 2.625 GB of user available memory and a memory bandwidth of 148.4 GB/sec.

Two private VMs \texttt{vm1} and \texttt{vm2} are employed. \texttt{vm1} is backed by Intel(R) Xeon(R) E5-2620 which is relatively old hardware when compared to the underlying processor for Amazon instances. \texttt{vm2} is backed by Intel(R) Xeon(R) E5-2665 similar to the underlying CPU on the \texttt{m2} instances. \texttt{vm2} is also supported by the same GPU on the Amazon GPU instance. The vCPUs of both VMs are hardware hyperthreads on the underlying processor and the Xen hypervisor is used.

The Ubuntu 13.10 cloud image is used on all VMs. Sequential and parallel versions of Aggregate Risk Analysis were implemented. C++ was used for the sequential implementation, OpenMP was used in the parallel implementation on multiple core instances, and CUDA was used for the sequential and parallel implementations on GPU instances. On the CPU instances, both versions were compiled using the GNU Compiler Collection g++ 4.6 and optimised using `\texttt{-O3}'; the parallel implementation required \texttt{-fopenmp} during compilation for including the OpenMP directive. The NVIDIA CUDA Compiler (nvcc 5.0) was used for compiling on the GPU instances. Message Passing Interface (MPI) was employed and added nearly 10\% to the execution times and was also found to lower the efficiency of the VMs; hence those results are not presented. However, for a multiple GPU instance a combination of MPI to communicate between the instances, OpenMP for exploiting parallelism of the virtual cores on each instance and CUDA programming for exploiting the GPU were employed.  

\begin{table}
\begin{center}
	\begin{tabular}{p{1.5cm} p{0.8cm} p{0.8cm} p{2.4cm} p{0.8cm} }
		\hline	
		\textbf{Instance Type}	&	\textbf{No. of Virtual CPUs (VCPU)}	&	\textbf{Memory (GiB)}	&	\textbf{Processor Type}	& \textbf{Clock Speed (GHz)}\\
		\hline
		\multicolumn{5}{c}{\emph{Amazon cloud VMs}}\\
		\hline
		\texttt{m1.xlarge}	&	4	&	15.0	&	Intel Xeon E5-2650	&	2.00	\\
		\hline
		\texttt{m2.xlarge}	&	2	&	17.1	&	Intel Xeon E5-2665	&	2.40	\\
		\texttt{m2.2xlarge}	&	4	&	34.2	&	Intel Xeon E5-2665	&	2.40	\\
		\texttt{m2.4xlarge}	&	8	&	68.4	&	Intel Xeon E5-2665	&	2.40	\\
		\hline
		\texttt{m3.xlarge}	&	4	&	15.0	&	Intel Xeon E5-2670	&	2.60	\\
		\texttt{m3.2xlarge}	&	8	&	30.0	&	Intel Xeon E5-2670	&	2.60	\\
		\hline
		\texttt{cr1.4xlarge}	&	32	&	244.0	&	Intel Xeon E5-2670	&	2.60\\		
		\hline
		\texttt{cc1.4xlarge}	&	16	&	23.0	&	Intel Xeon X5570	&	2.93\\
		\texttt{cc2.8xlarge}	&	32	&	60.5	&	Intel Xeon X5570	&	2.93\\
		\hline		
		\texttt{hi1.4xlarge}	&	16	&	60.5	&	Intel Xeon E5620	&	2.40\\
		\texttt{hs1.4xlarge}	&	16	&	117.0	&	Intel Xeon E5-2650	&	2.00\\
		\hline
		\multirow{2}{*}{\texttt{cg1.4xlarge}}	&	16	&	22.5	&	Intel Xeon X5570	&	2.93	\\
								&	448	&	3.0		&	NVIDIA Tesla M2050	&	0.575  		\\
		\hline
		
		\multicolumn{5}{c}{\emph{Private VMs}}\\
		\hline
		\texttt{vm1}	&	12	&	128.0	&	Intel Xeon E5-2620	&	2.00	\\
		\hline
		\multirow{2}{*}{\texttt{vm2}}	&	16	&	64.0	&	Intel Xeon E5-2665	&	2.40	\\
										&	448	&	3.0		&	NVIDIA Tesla M2050	&	0.575  	\\
		\hline
	\end{tabular}
	\caption{VMs employed for Aggregate Risk Analysis}
	\label{table1}
	\end{center}
\end{table}

\subsection{Implementation on the Cloud}
In all the implementations each trial in the YET is executed using a single thread. All data required for the analysis is available as an Amazon Elastic Block Storage (EBS)\footnote{\url{http://aws.amazon.com/ebs/}} volume which is attached onto an instance. Nearly fifteen hours of continuous data transfer was required to the EBS volume. 

\begin{figure}[t]
\centering
	\subfloat[Time taken on VMs]{\label{graphset0-1}\includegraphics[width=0.49\textwidth]{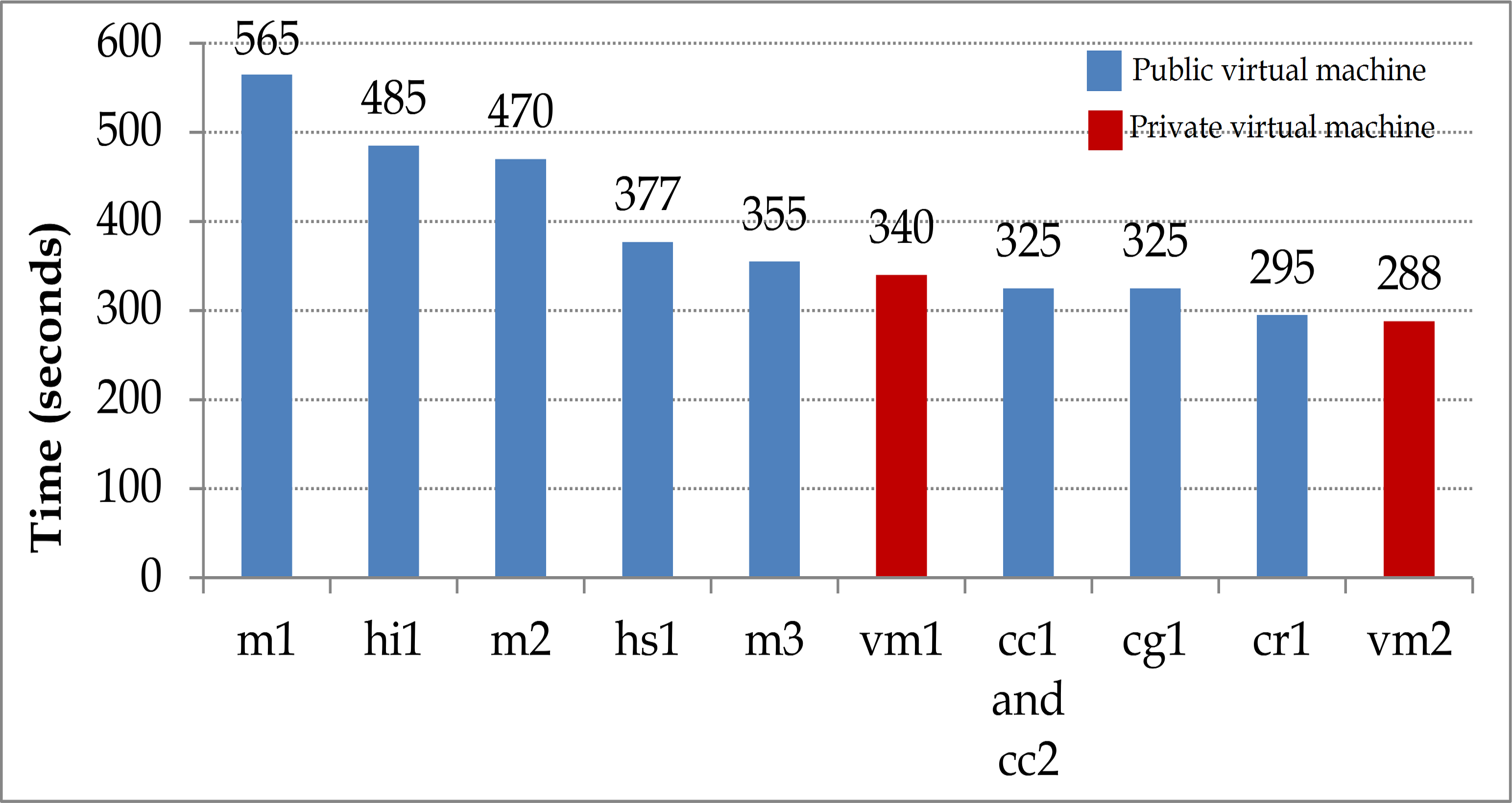}} \\%\hspace{12pt}
	\subfloat[Time taken when number of Trials are varied on the fastest public and private VMs]{\label{graphset0-2}\includegraphics[width=0.49\textwidth]{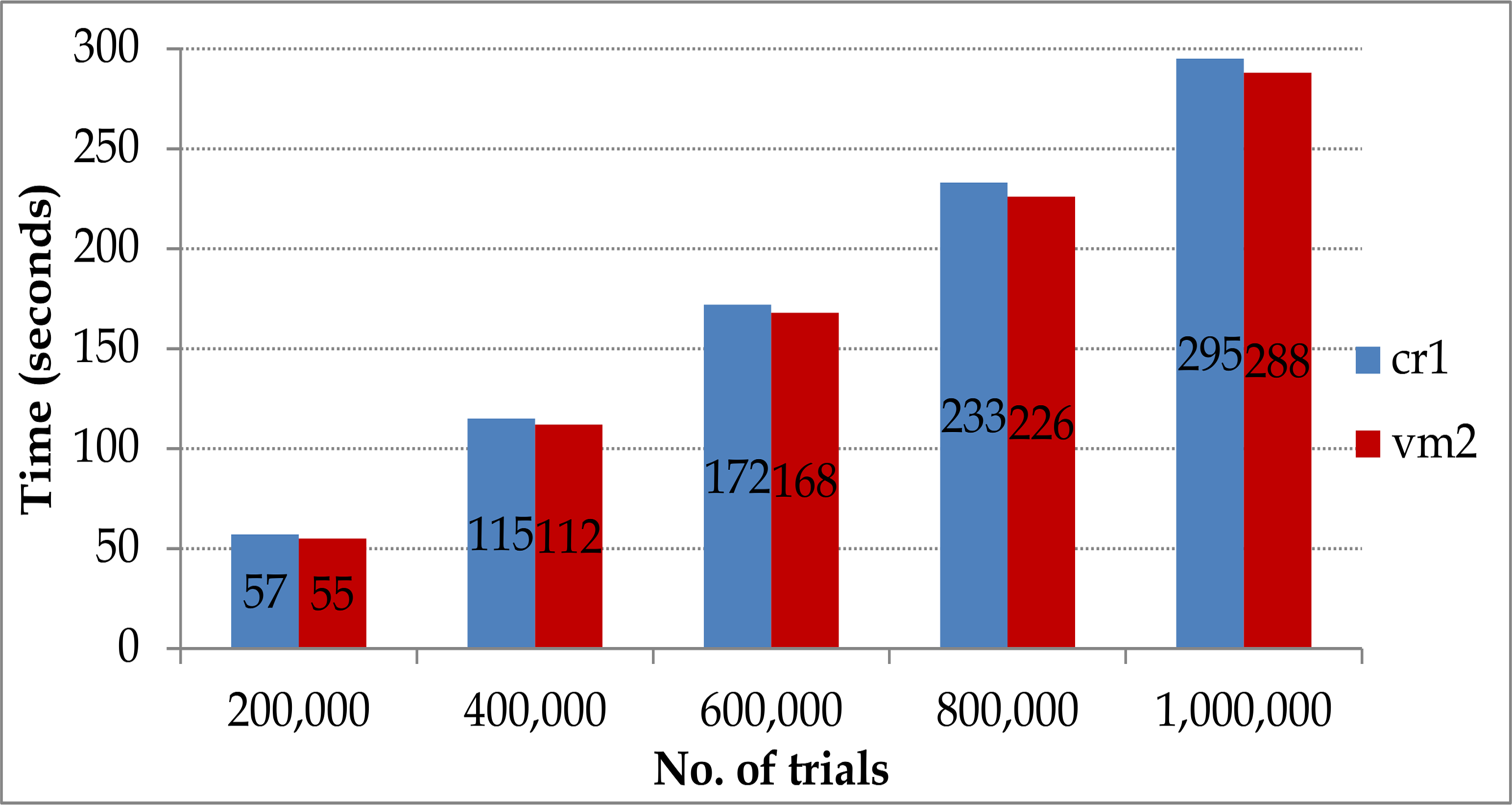}}
\caption{Sequential performance of Aggregate Risk Analysis on VMs}
\label{graphset0}
\end{figure}

One important factor for obtaining good performance in the Aggregate Risk Analysis algorithm is the selection of a data structure for representing ELTs. The ELTs function as dictionaries with key-value pairs requiring fast random key lookup. A sparse representation of ELTs covered by a Layer using direct access tables was implemented. Although fast lookups are obtained a sparse representation is at the expense of high memory utilisation. If a YET consists of 1,000,000 events and an ELT consists of 10,000 event-loss pairs, then the direct access table would contain 1,000,000 event entries of which 990,000 events would have zero loss values. Considering that one Layer would cover fifteen ELTs in a typical analysis, 15 million event-loss pairs need to be generated in the memory of the instance of which 14,850,000 events have zero loss values. 

Nevertheless a direct access table was employed in all implementations. Alternate compact representations were not chosen for the following reasons: (i) A search operation is required to find an event-loss pair in a compact representation. Sequential and binary search require $O(n)$ and $O(log(n))$ memory accesses respectively to locate an event-loss pair. Even if a constant-time space-efficient hashing scheme requiring a constant number of memory accesses is adopted there is considerable implementation and run-time performance complexity. This overhead will be high on GPUs with complex memory hierarchies consisting of global and shared memories. (ii) To perform aggregate analysis on a YET of one million trials, each trial comprising one thousand events, and for a layer covering fifteen ELTs, there are fifteen billion event-loss pairs. Direct access tables, although require large memory space, allow for the least number of memory accesses as each lookup in an ELT requires only one memory access per search operation.

Two data structure implementations of the ELTs were considered. In the first implementation, each ELT is an independent table, and therefore, in a read cycle, each thread independently looks up its events from the ELTs. All threads within a block access the same ELT. By contrast, in the second implementation, the ELTs are combined as a single table. Consequently, the threads then use the shared memory to load entire rows of the combined ELTs at a time. The latter implementation performs poorly compared to the former since for the threads to collectively load from the combined ELT each thread must first write which event it requires. This results in additional memory overheads. 

On the CPU instance offering multiple virtual cores the entire data required for the analysis is processed in memory. The GPU implementation uses the GPU's global memory to store all of the required data structures. The parallel implementation on the GPU requires high memory transactions and leads to inefficient performance on the GPU platform. To surmount this challenge shared memory can be utilised over global memory. 

The algorithm is optimised in the following four ways. Firstly, by chunking, which refers to processing a block of events of fixed size (or chunk size) for the efficient use of shared memory. 

Chunking is more beneficial in the GPU implementation than in the CPU implementations. In the case of the GPU implementation looking up events in a trial and applying financial terms to losses at the Event and Layer level are chunked. Further, the financial and layer terms are stored in the streaming multi-processor's constant memory. 

If the intermediate losses are represented in global memory, then while applying the financial terms at the Event and Layer level would require the global memory to be accessed and updated adding considerable memory overheads. The memory overhead is minimised by chunking when (i) the financial terms are applied, and (ii) reading events in a trial from the YET. Chunking reduces the number of global memory update and global read operations. Moreover, the benefits of data striding can also be used to improve speed-up. 

Secondly, the implementation are optimised by loop unrolling, which refers to the replication of blocks of code included within for loops by the compiler to reduce the number of iterations performed by the for loop. This is done using the pragma directive. 

Thirdly, the implementations on the CPU and GPU are optimised by using single precision operations when possible. Read operations are faster using float variables as they are only half the size of a double variable. The performance of single precision operations tend to be approximately twice as fast as double precision operations.

Fourthly, in the case of the GPU a further optimisation can be achieved by migrating data from both shared and global memory to the kernel registry. The kernel registry has the lowest latency compared to all other forms of memory available in the GPU architecture.

\subsection{Empirical Analysis}
\begin{figure*}
\centering
	\subfloat[Two virtual core public instance]{\label{graphset1-1}\includegraphics[width=0.33\textwidth]{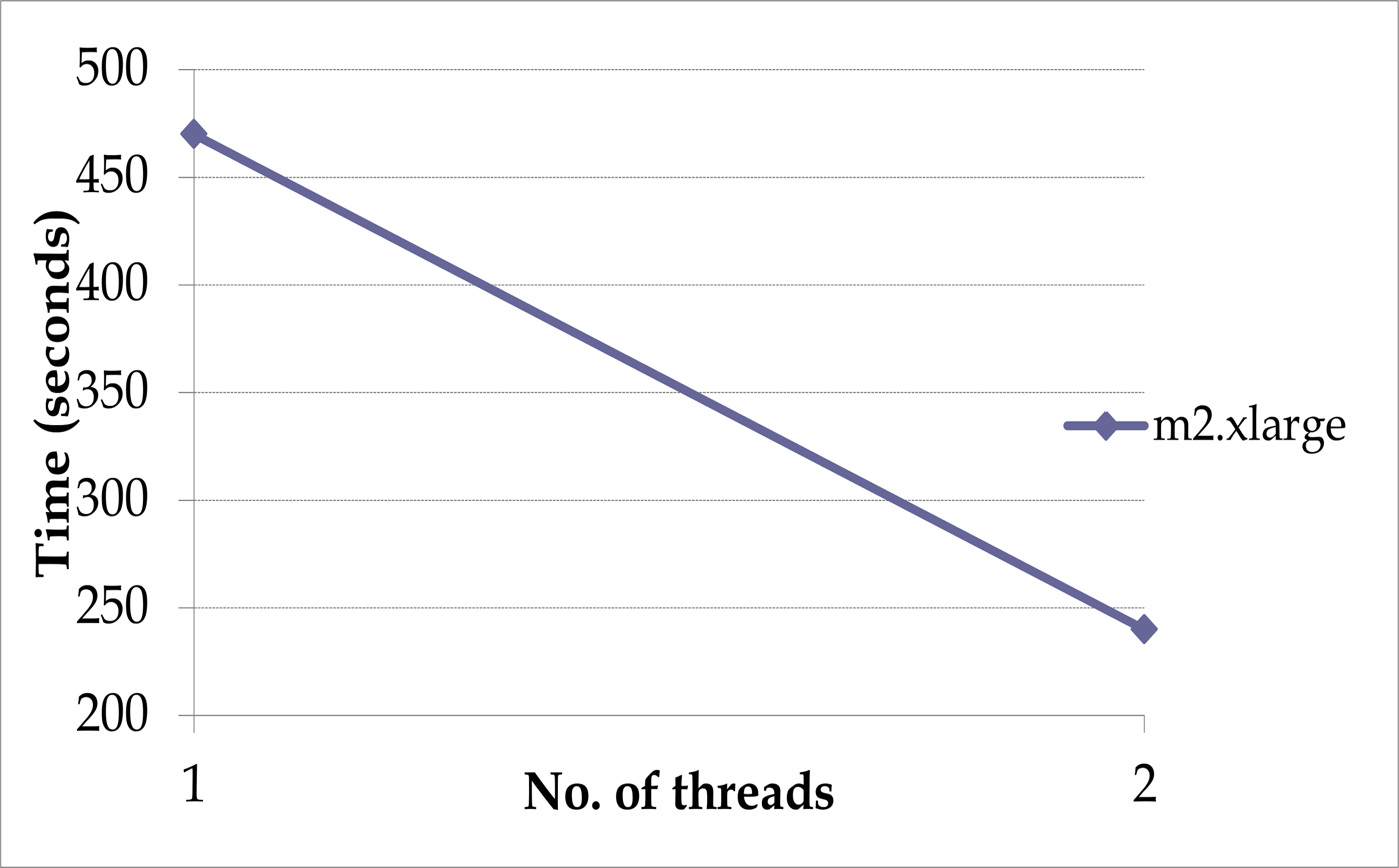}} \hfill
	\subfloat[Four virtual core public instances]{\label{graphset1-2}\includegraphics[width=0.33\textwidth]{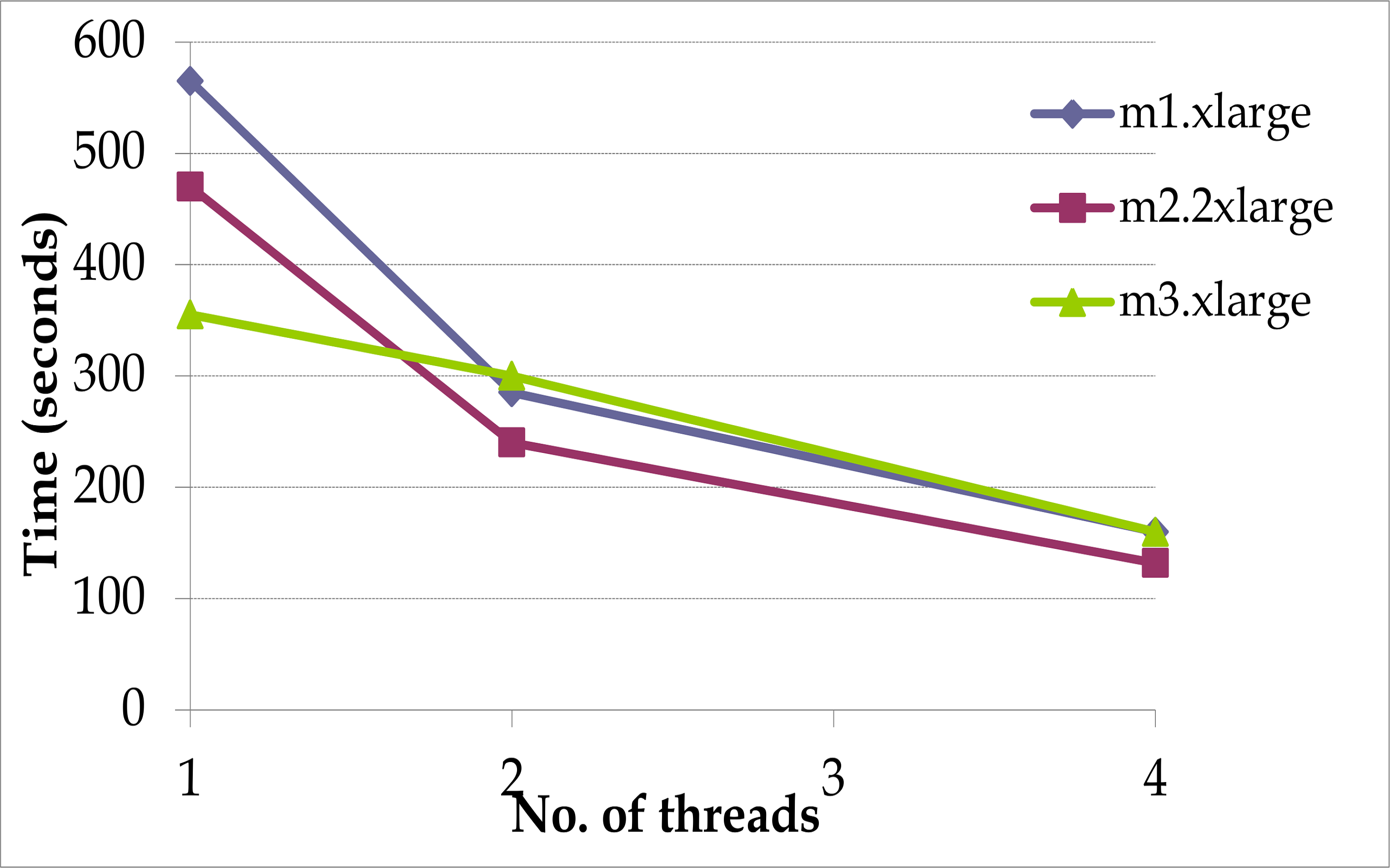}} \hfill
	\subfloat[Eight virtual core public instances]{\label{graphset1-3}\includegraphics[width=0.33\textwidth]{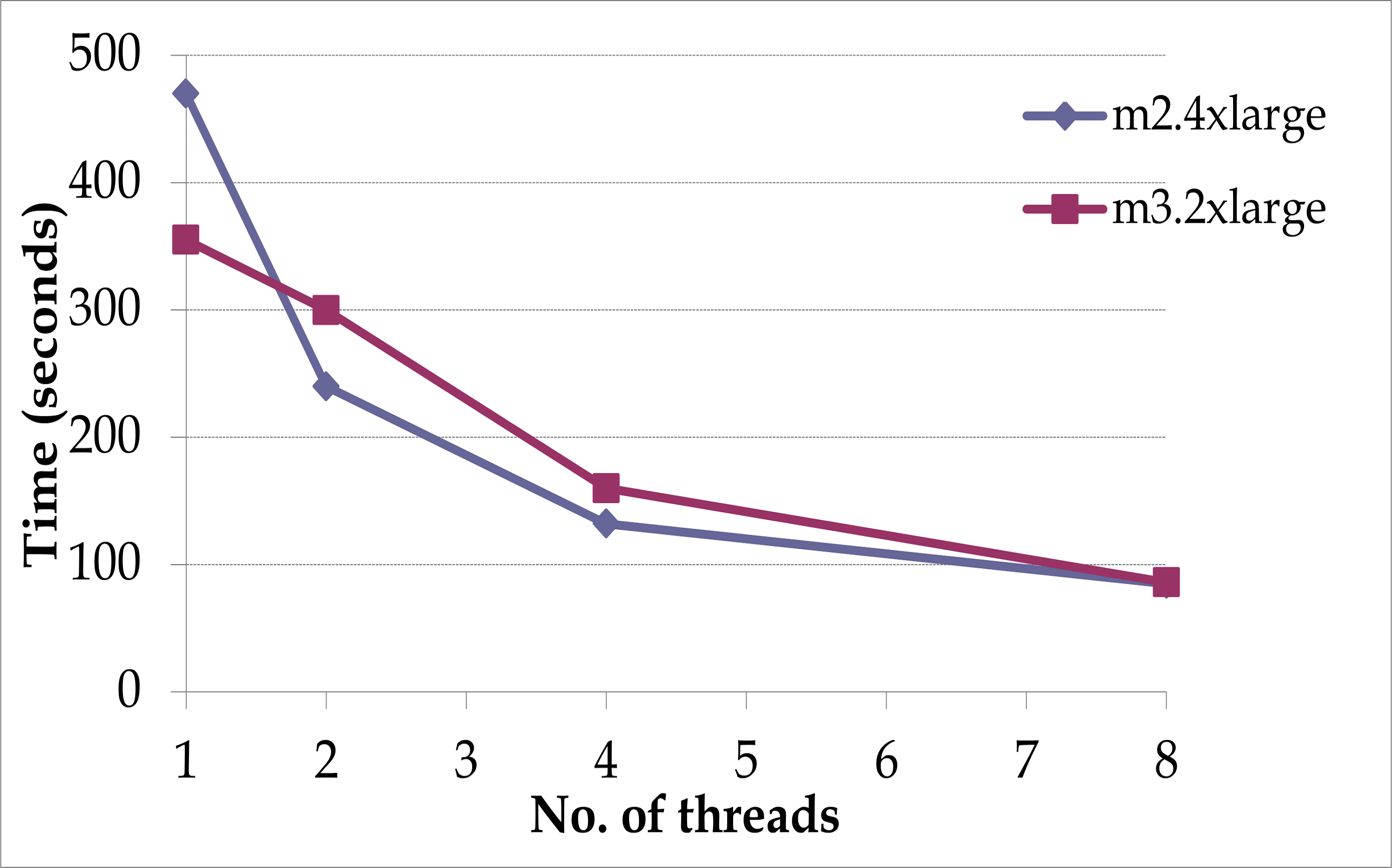}} \\
	\subfloat[Sixteen virtual core public instances]{\label{graphset1-4}\includegraphics[width=0.33\textwidth]{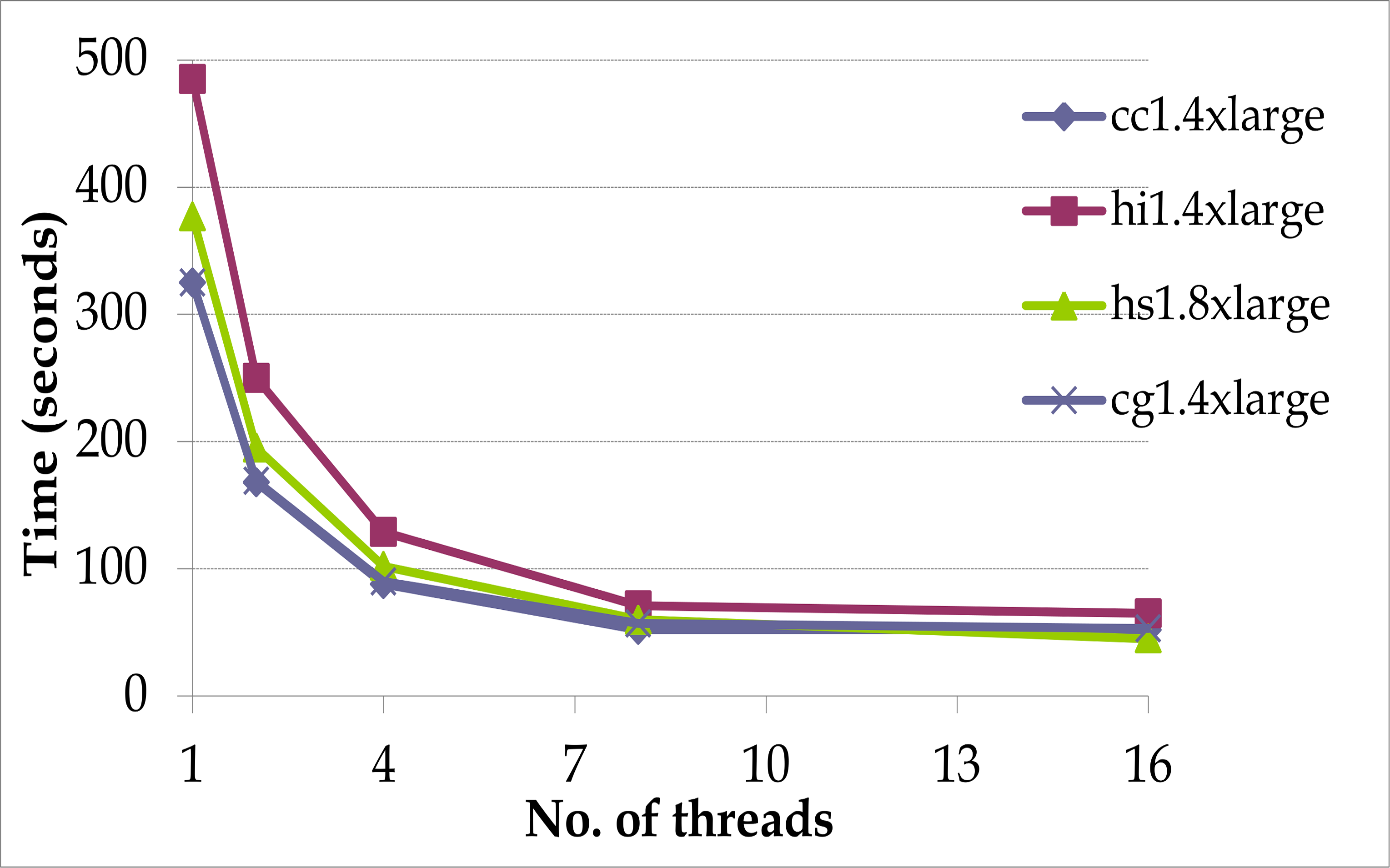}} \hfill
	\subfloat[Thirty two virtual core public instances]{\label{graphset1-5}\includegraphics[width=0.33\textwidth]{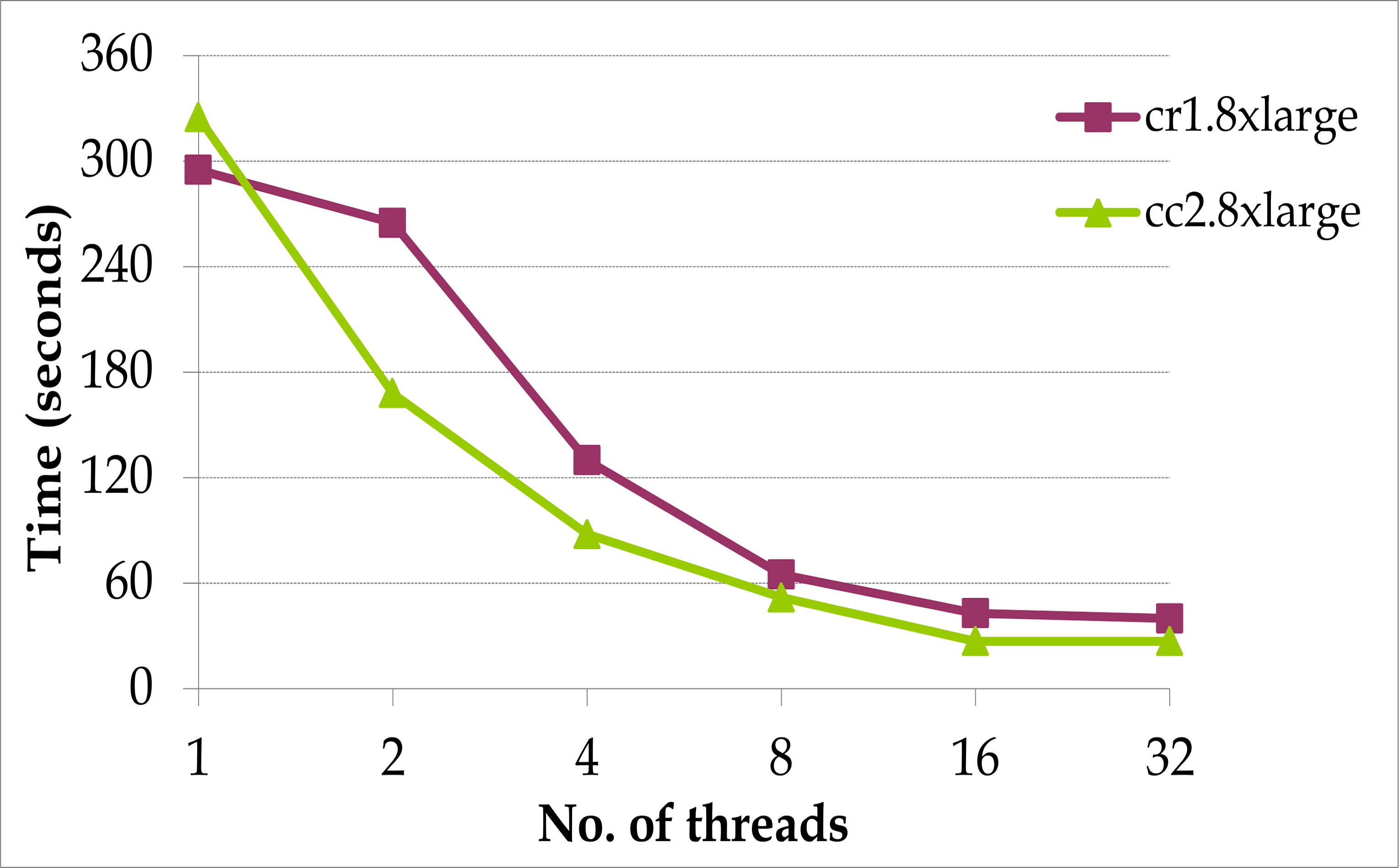}} \hfill
	\subfloat[Private VMs]{\label{graphset1-6}\includegraphics[width=0.33\textwidth]{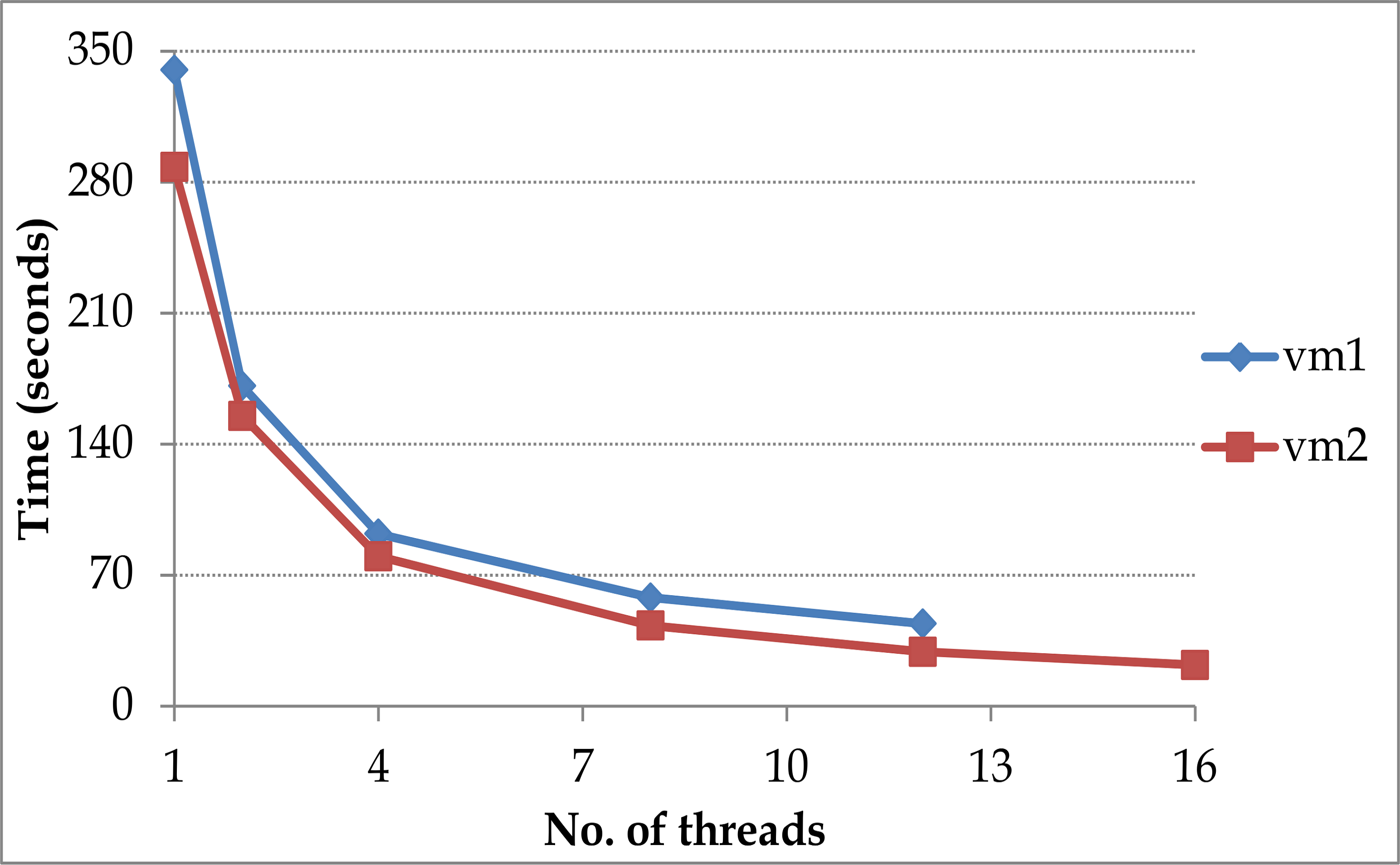}} 
\caption{Parallel execution (single thread per virtual core) of Aggregate Risk Analysis on public (a) to (e) and private (f) VMs}
\label{graphset1}
\end{figure*}

The results obtained from the experimental studies are presented in this section. All data required for the analysis is stored as an EBS volume and attached onto the instances considered in Table \ref{table1}. Figure \ref{graphset0} to Figure \ref{graphset3} are results obtained on CPU instances; the multi-core architecture of the instances are exploited in the parallel implementation. Figure \ref{graph4} and Figure \ref{graphset5} are results obtained on the GPU instance; both single and multiple GPUs are exploited in the parallel implementation. In all experiments, the analysis uses as input a YET comprising one million trials, with each trial consisting of one thousand catastrophic events, and one Portfolio with one Program comprising one Layer covering sixteen ELTs. The input parameters are realistic and were chosen based on industry-wide practices. 

\subsubsection{Results from CPU instances}
Figure \ref{graphset0-1} shows the graph plotted for the time taken for sequentially performing aggregate risk analysis on all instances shown in Table \ref{table1}. Under the general purpose instances, the \texttt{m1} is the slowest for performing the analysis requiring 565 seconds; the \texttt{m3} instance is over 37\% faster than the \texttt{m1} instance. The memory optimised instance \texttt{cr1} is the fastest for performing the analysis requiring 295 seconds which is 37\% faster than the memory optimised \texttt{m2} instances. The difference in the performance obtained on storage optimised instances is just over 20\%. Cluster instances \texttt{cc1} and \texttt{cc2} perform comparably to the \texttt{cg1} instance. The fastest sequential CPU performance on the cloud requires less than five minutes which is nearly 50\% fastest than the slowest sequential performance on the cloud. Private VMs \texttt{vm1} and \texttt{vm2} have surprisingly good performance. \texttt{vm1} takes only 340 seconds which is nearly 40\% faster than \texttt{m1} instances. The \texttt{vm2} VM completes the analysis in 288 seconds which is over 2\% faster than the best performance on Amazon. Figure \ref{graphset0-2} shows the both the increase in the total time taken for the fastest sequential analysis on the cloud when the number of trials are varied between two hundred thousand and one million trials.

The parallel implementation of the analysis on the CPU requires multiple threads to be executed on the instance which can be done in two ways. Firstly, by executing a single thread per virtual core, and secondly, by executing multiple threads per core. 

Figure \ref{graphset1} shows the graphs obtained from the parallel implementation of the analysis when one thread is executed per virtual core on the instance. The graphs are organised based on the number of virtual cores on the instance. The instance with two virtual cores obtains nearly a 96\% efficiency when two threads are employed (Figure \ref{graphset1-1}). Instances with four virtual cores obtain upto 87.5\% efficiency (Figure \ref{graphset1-2}). The two instances with eight virtual cores have an average efficiency of over 70\% (Figure \ref{graphset1-3}). 

The storage optimised, \texttt{cc1} and \texttt{cg1} instances with sixteen cores each exhibit very little speedup and efficiency for more than eight cores (Figure \ref{graphset1-4}). Surprisingly, there is no hardware acceleration obtained which is expected. Beyond eight cores it would seem that the cost of hardware and the use of virtualised hardware on the sixteen core VMs do not benefit the analysis. Another reason is that as the number of cores are increased the bandwidth to access memory is not equally increased which is a limiting factor. 

Similarly, in the case of thirty two core instances no acceleration is obtained beyond sixteen cores (Figure \ref{graphset1-5}). The fastest parallel execution on the CPU is obtained on the cluster compute instance \texttt{cc2.8xlarge} taking 27 seconds with a speedup of nearly 11x over the fastest sequential implementation. The performance of \texttt{cr1.8xlarge} is second to the \texttt{cc2} instance requring 40 seconds when multiple threads are employed though it performs the sequential analysis the fastest. 

The private VMs again outperform the public instances (Figure \ref{graphset1-6}). \texttt{vm1} takes 44 seconds achieving a speedup of 7.5x over its sequential performance and \texttt{vm2} takes 22 seconds achieving a speedup of 13x over its sequential performance. 

\begin{figure*}
\centering
	\subfloat[Two virtual core public instance]{\label{graphset2-1}\includegraphics[width=0.33\textwidth]{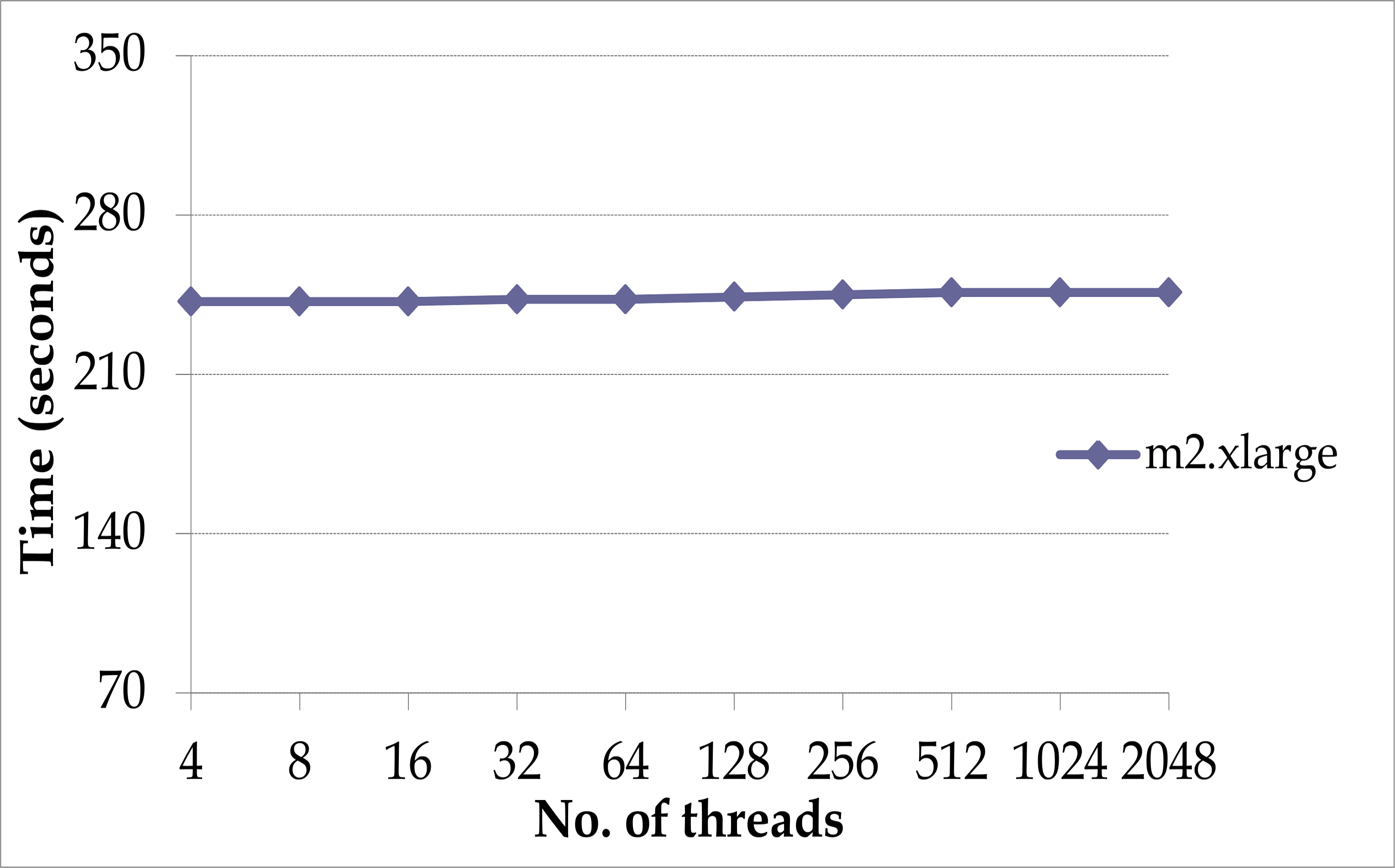}} \hfill
	\subfloat[Four virtual core public instances]{\label{graphset2-2}\includegraphics[width=0.33\textwidth]{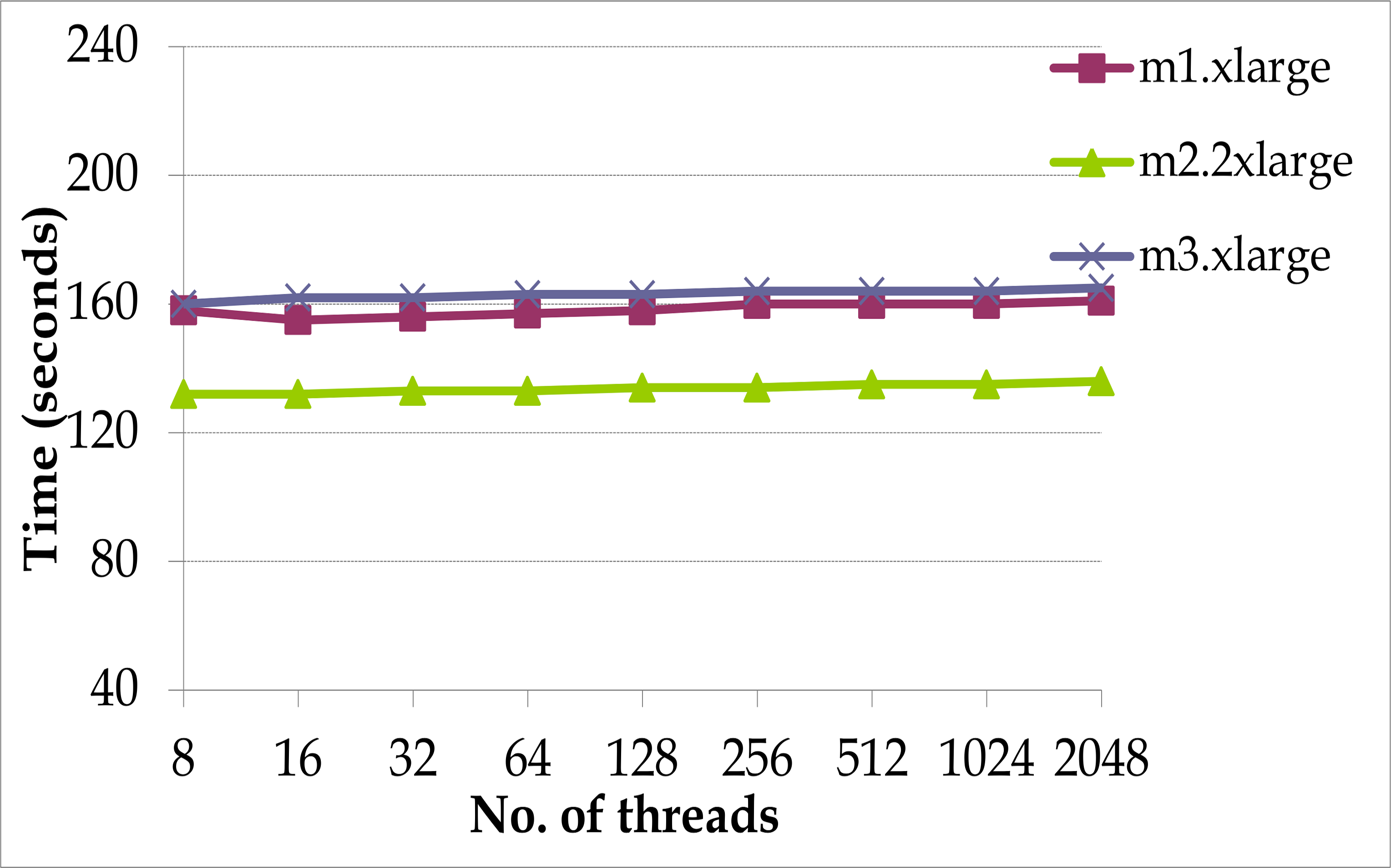}} \hfill
	\subfloat[Eight virtual core public instances]{\label{graphset2-3}\includegraphics[width=0.33\textwidth]{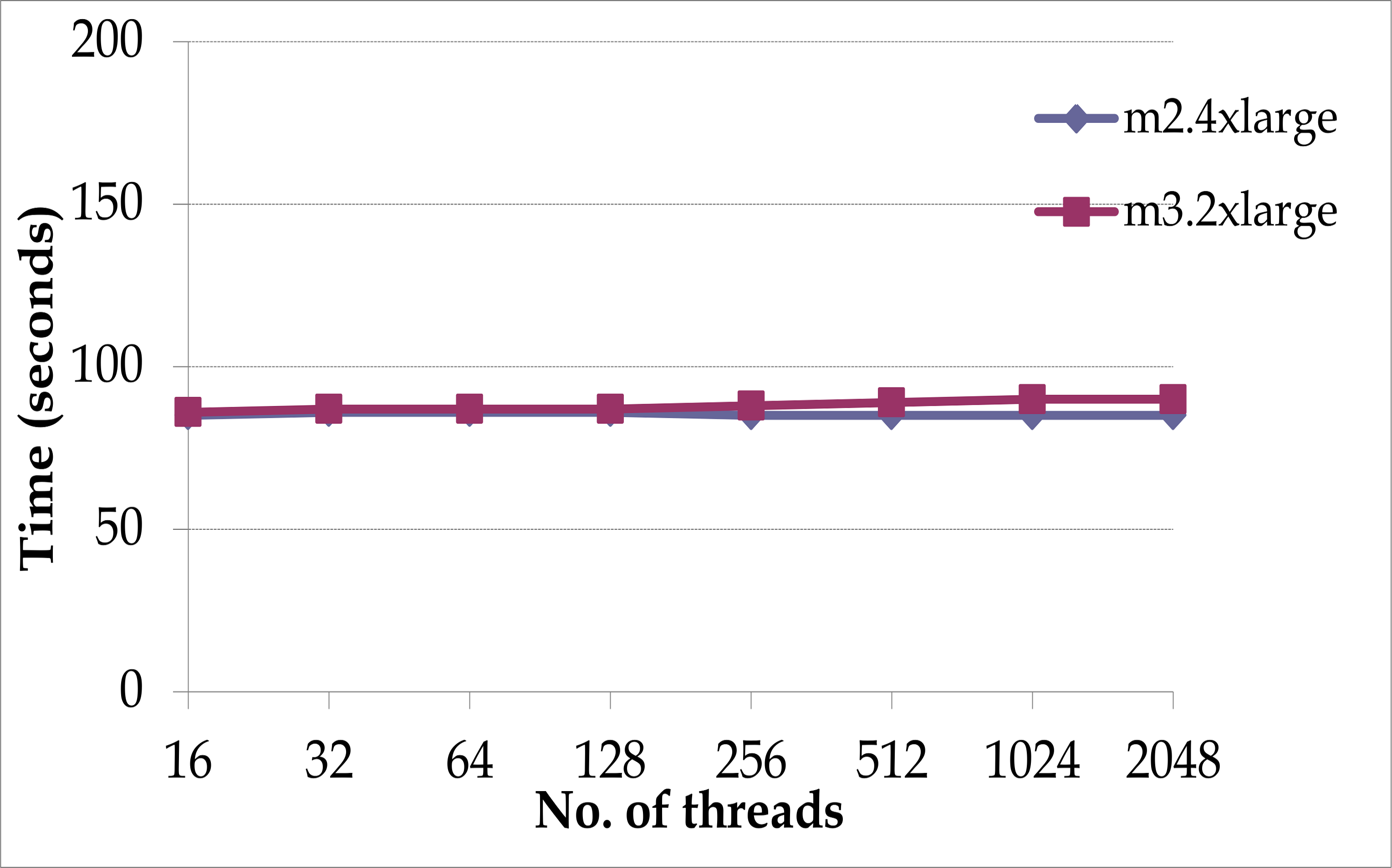}} \\
	\subfloat[Sixteen virtual core public instances]{\label{graphset2-4}\includegraphics[width=0.33\textwidth]{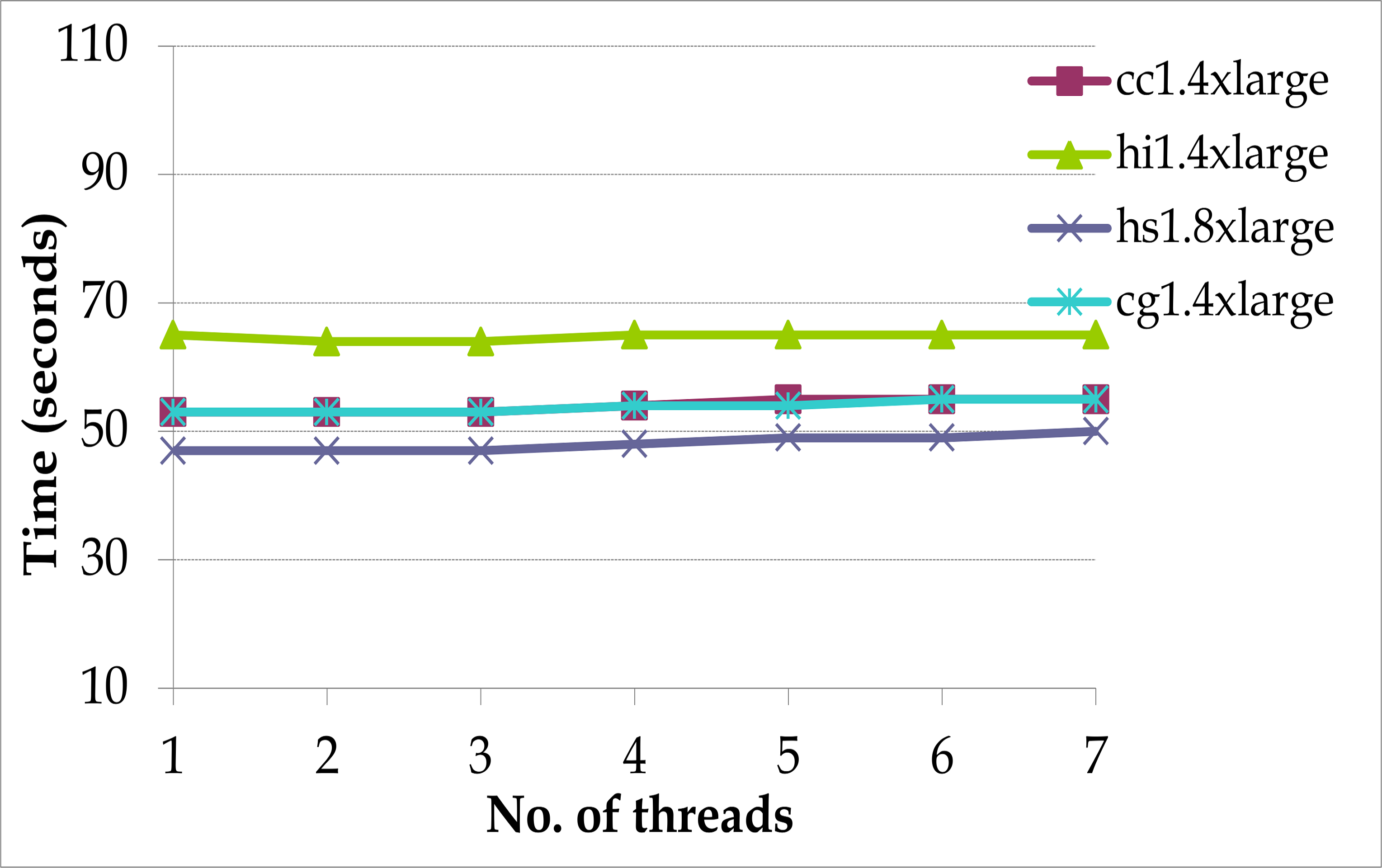}} \hfill
	\subfloat[Thirty two virtual core public instances]{\label{graphset2-5}\includegraphics[width=0.33\textwidth]{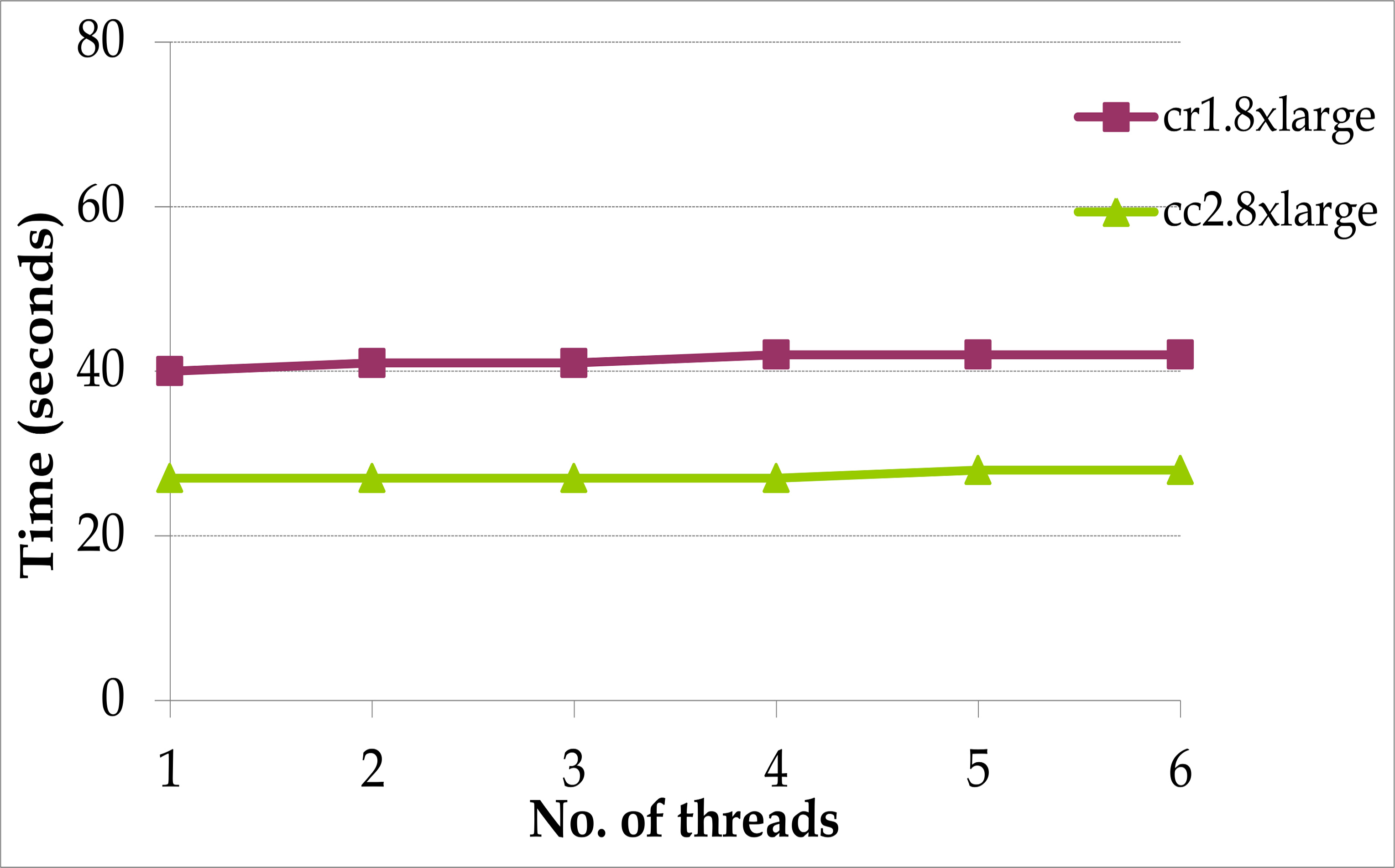}} \hfill
	\subfloat[Private VMs]{\label{graphset2-6}\includegraphics[width=0.33\textwidth]{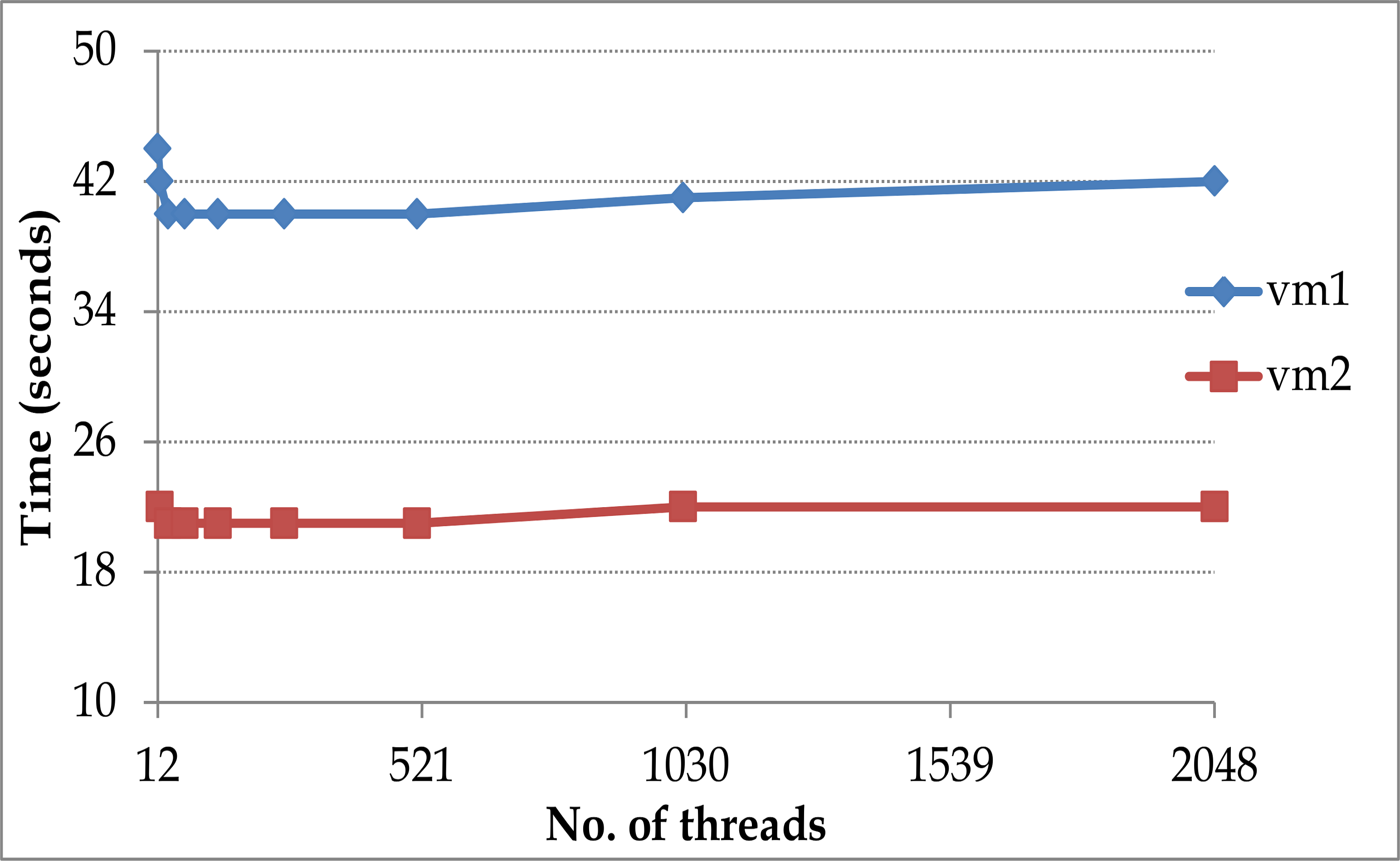}}
\caption{Parallel execution (multiple threads per virtual core) of Aggregate Risk Analysis on public (a) to (e) and private (f) VMs}
\label{graphset2}
\end{figure*}

Figure \ref{graphset2} shows the graphs obtained from the parallel implementation of the analysis when multiple threads are executed on the instances. In all cases, multiple threads per Amazon core do not provide any acceleration for the analysis. Increasing the number of threads per core results in an increase in the communication cost between threads. The private VMs \texttt{vm1} and \texttt{vm2} achieve a speedup of 9\% and 5\% respectively when multiple threads are employed per virtual core. 

\begin{figure}[t]
\centering
	\subfloat[Time taken on VMs]{\label{graphset3-1}\includegraphics[width=0.49\textwidth]{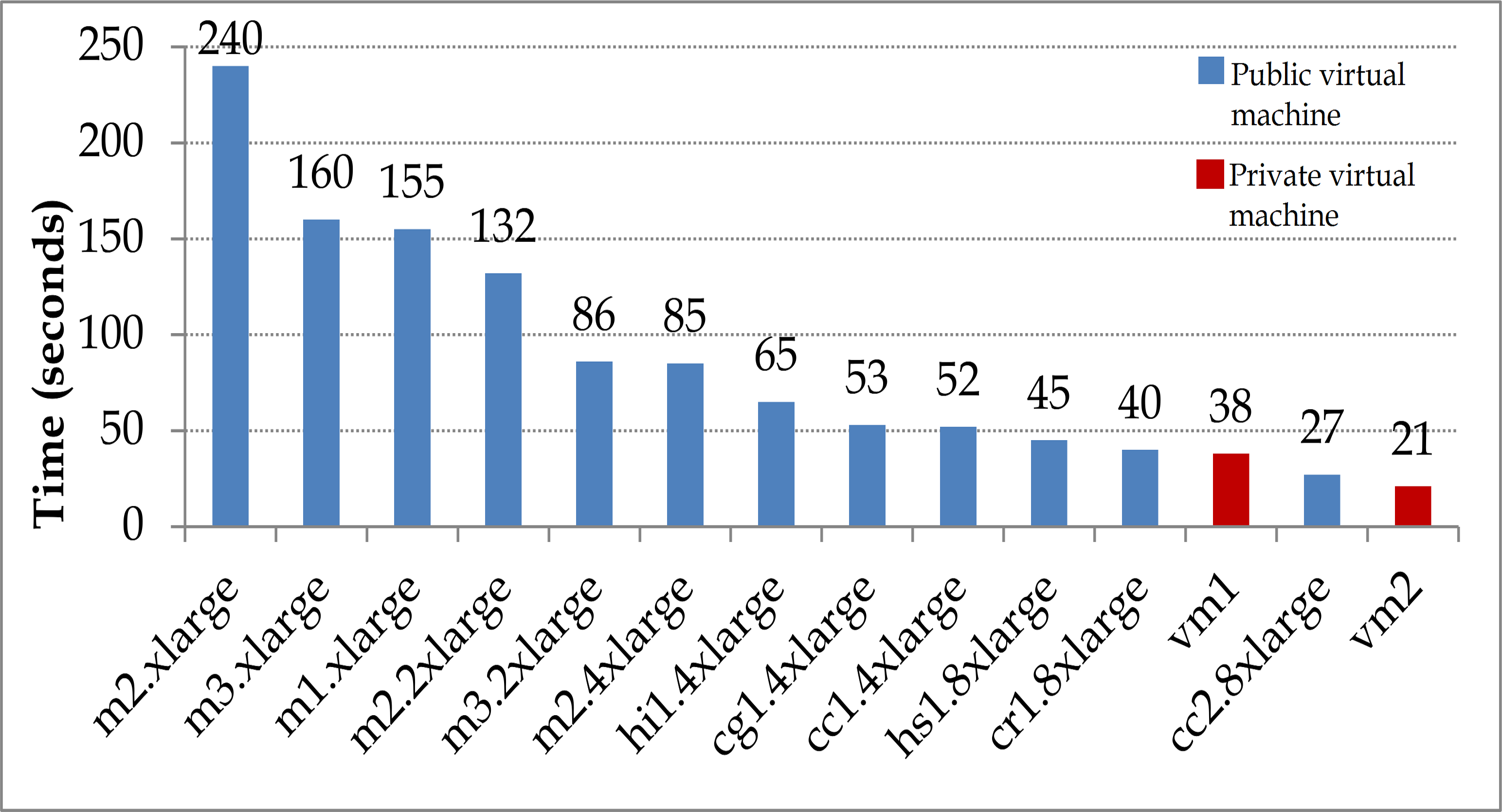}}\\ %\hspace{12pt}
	\subfloat[Time taken when number of Trials are varied on the fastest public and private VMs]{\label{graphset3-2}\includegraphics[width=0.49\textwidth]{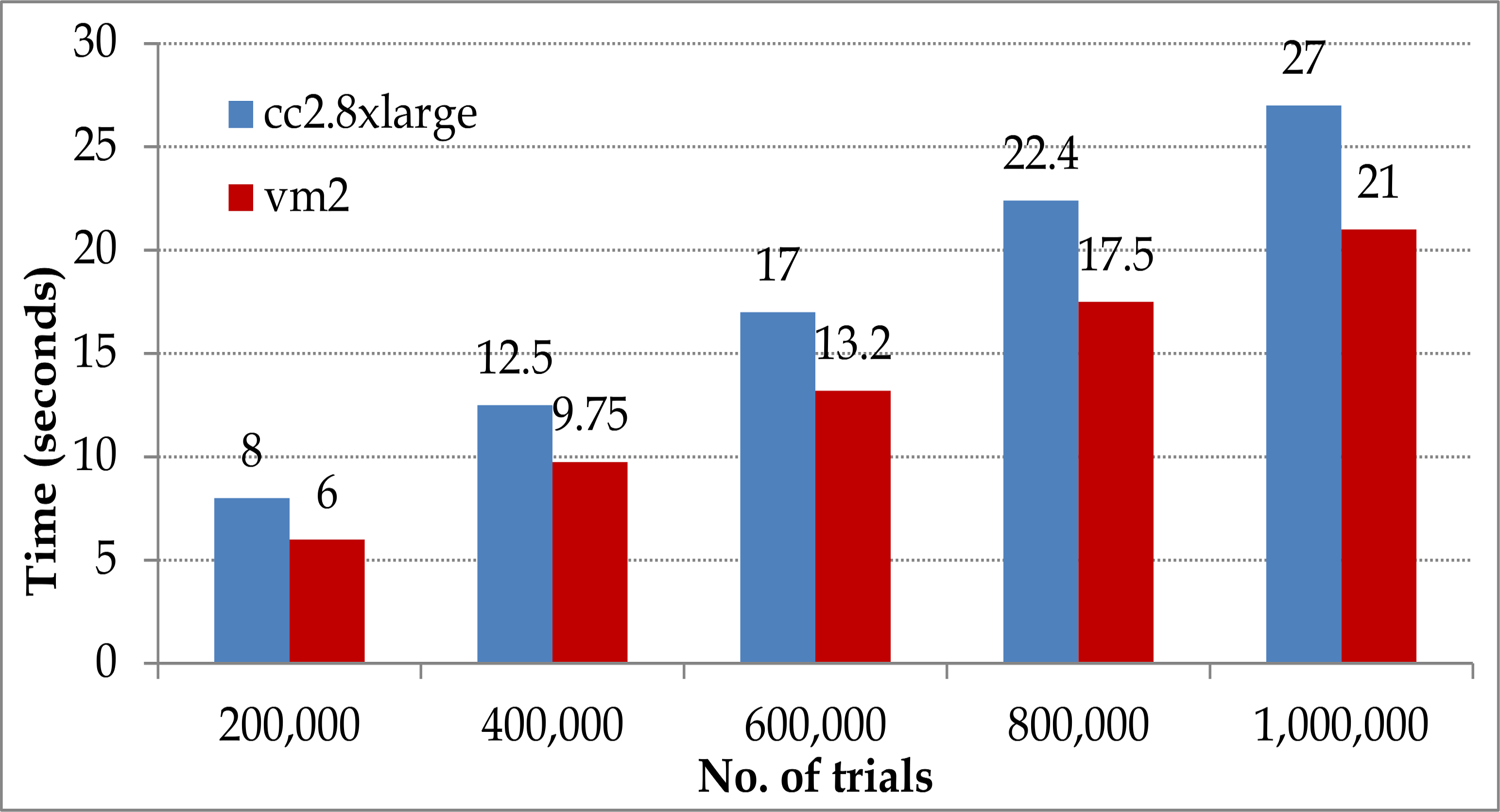}} \\
\caption{Parallel execution of Aggregate Risk Analysis on public and private VMs}
\label{graphset3}
\end{figure}

Figure \ref{graphset3-1} shows the graph plotted for the best time taken for performing parallel aggregate risk analysis on all instances shown in Table \ref{table1}. Under the general purpose instances, though the \texttt{m1} instance is the slowest for performing the sequential analysis and for parallel analysis the \texttt{m2.xlarge} is the slowest requiring 240 seconds. Virtual core acceleration is achieved on the \texttt{m1} instance which is over 1.5x faster than \texttt{m2.xlarge}. The \texttt{m3.2xlarge} is nearly 2 times faster than \texttt{m3.xlarge} and \texttt{m1.xlarge}. The cluster instance \texttt{cc2} followed by \texttt{cr1} are the fastest requiring 27 seconds and 40 seconds respectively. Hence upto a 21x speedup is obtained for parallel analysis by exploiting the multi-core architecture over the sequential analysis and upto a 9x speedup over the slowest parallel analysis. Again, private VMs outperform public instances. The best performance of of \texttt{vm2} is 21 seconds using multiple threads which is 22\% faster than the best performance achieved by the \texttt{cc2} instance. Similarly, \texttt{vm1} takes 38 seconds for the analysis on multiple threads which is 5\% faster than the second best performance by \texttt{cr1} instances on public instances. 

Figure \ref{graphset3-2} shows the increase in the total time taken for the fastest parallel analysis on the cloud when the number of trials are varied between two hundred thousand and one million trials. 

\subsubsection{Results from GPU instance}
Single and multiple GPU instances (\texttt{cg1.4xlarge}) are considered for risk analysis on the cloud. CUDA provides an abstraction over the streaming multi-processors of the GPU, referred to as a CUDA block. Unlike the parallel implementations on the CPU instance an additional parameter that needs to be considered in the GPU implementations is the number of threads executed per CUDA block. To represent 1,000,000 trials of the analysis on the GPU instance consider each trial is executed on one thread. If 128 threads are executed on one streaming multi-processor there will be approximately 7813 blocks which need to be executed on 14 streaming multi-processors; each streaming multi-processor will therefore need to execute 558 blocks. All threads executed on a streaming multi-processor share fixed allocations of shared and constant memory. Therefore, there is a trade-off for optimal performance; each thread can access larger amounts of shared and constant memory if there are fewer threads per block, but then the global memory will required to be accessed more resulting in increased global memory overheads. 

\begin{figure}
\centering
	\includegraphics[width=0.49\textwidth]{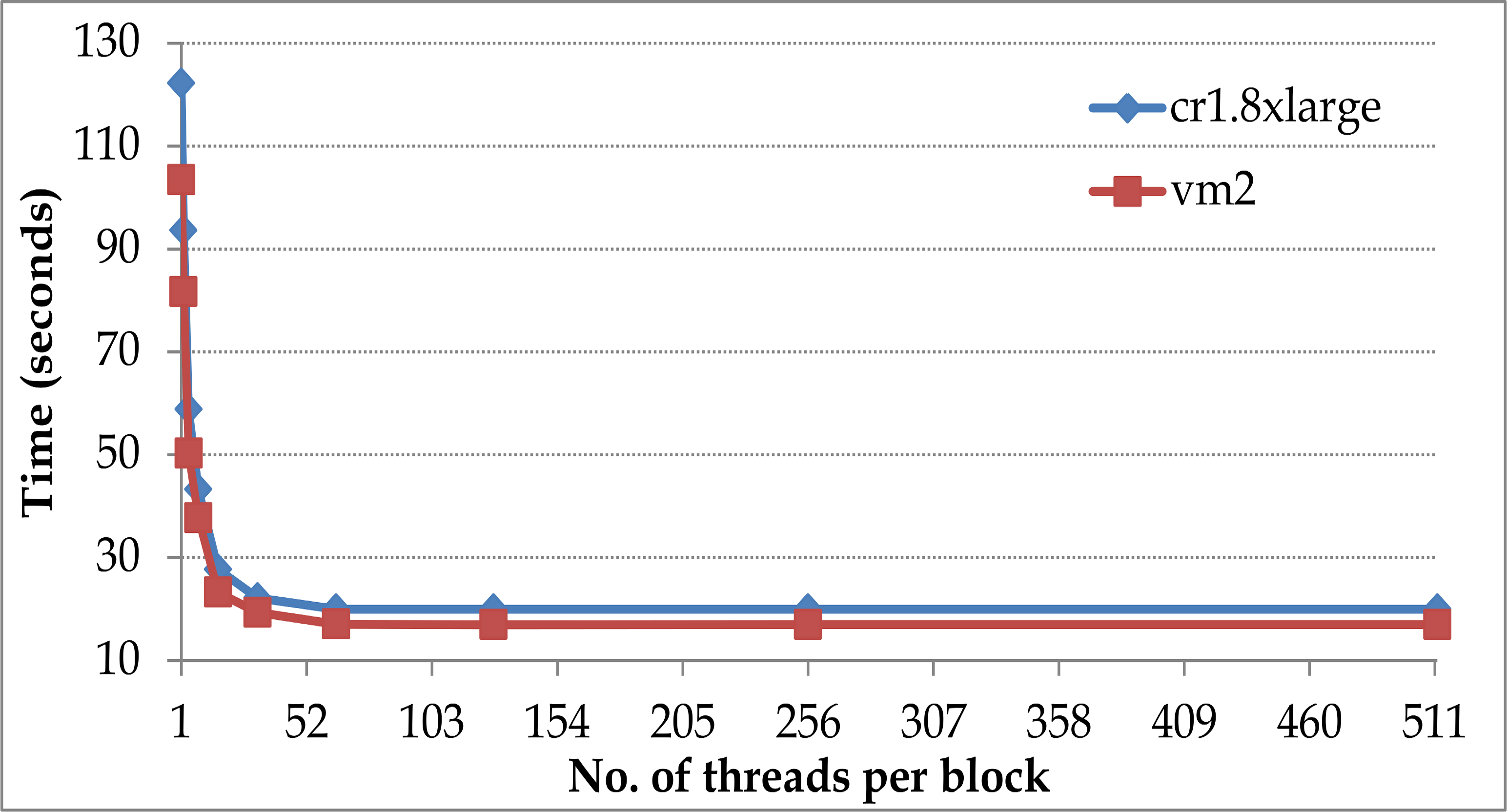}
	\caption{Execution of Aggregate Risk Analysis on a single GPU on public and private VMs}
	\label{graph4}
\end{figure} 

Figure \ref{graph4} shows the time taken for parallel risk analysis on a single GPU instance when the number of threads per CUDA block are varied between 1 and 512. At least 16 threads per block are required to achieve performance comparable to the best parallel implementation which is noted on \texttt{cc2.8xlarge} instance. To exploit the full potential of hardware acceleration offered by GPUs at least 64 threads are required. An improvement in the performance is observed with 256 threads per block beyond which performance starts to diminish. The best time for performing the analysis on a single GPU is 19.93 seconds which is around 15x faster than the best sequential performance on the CPU, nearly 1.4x faster than the best multiple core performance on the CPU, and over 6x faster than the sequential performance on a GPU. On the private VM \texttt{vm2} takes only 16.86 seconds. This is nearly 16\% faster than the GPU on the public instance. 

\begin{figure}
\centering
	\subfloat[Performance of multiple GPUs]{\label{graphset5-1}\includegraphics[width=0.49\textwidth]{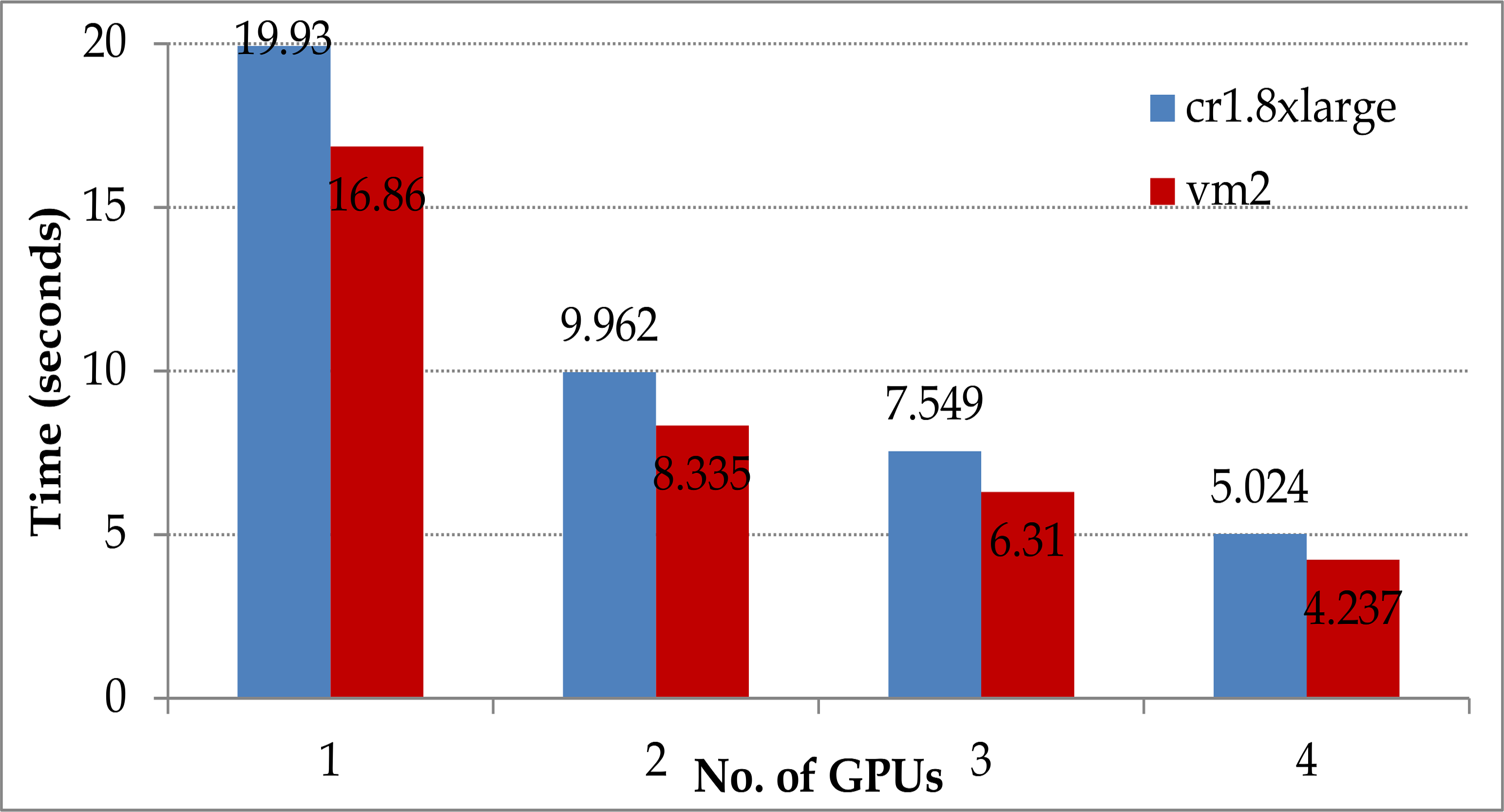}} \hspace{12pt}
	\subfloat[Performance on four GPUs for varying threads per block]{\label{graphset5-2}\includegraphics[width=0.49\textwidth]{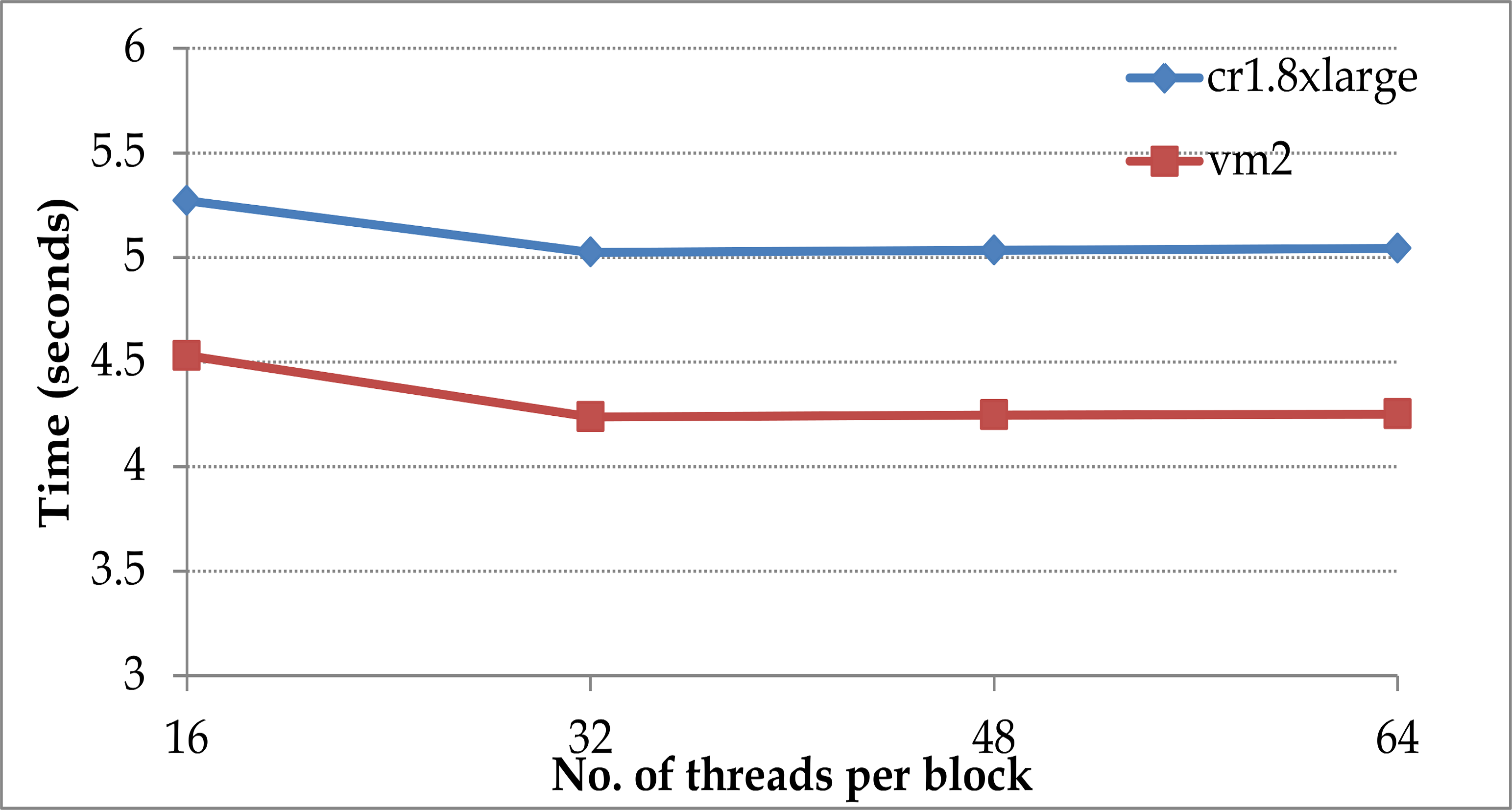}} \\
\caption{Aggregate Risk Analysis on public and private VMs with multiple GPUs}
\label{graphset5}
\end{figure}

The performance of the analysis on multiple GPU instances is shown in Figure \ref{graphset5-1}. In the multiple GPU implementation the workload for the analysis is decomposed and made available to the multiple instances that employ GPUs. Each CPU thread schedules the workload on the GPUs. Time taken for the analysis on four Amazon GPU instances is 5.024 seconds which is approximately 3.97 times faster than employing a single GPU with close to 97\% efficiency. Compared to the sequential implementation on a single GPU a speedup of over 24x is achieved when multiple GPU instances are used. On the other hand \texttt{vm2} takes only 4.238 seconds which is 16\% faster than the multiple GPUs on the public instance. 

Figure \ref{graphset5-2} shows the performance of the analysis on four GPUs when the number of threads per block is varied from 16 to 64. Experiments could not be pursued beyond 64 threads per block due to the limitation on the block size the shared memory can use. The best performance of 5.024 seconds on the public VM and 4.2375 seconds on the private VM is achieved when the number of threads per block is 32; the block size is the same as the WARP size of the GPU, therefore, an entire block of threads can be swapped when high latency operations occur. Increasing the number of threads per block does not improve the performance since there is shared memory overflow.

\subsection{Summary}
The experimental studies indicate that the public clouds are a suitable platform for accommodating risk simulations. The data, memory and computational requirements can be met on the cloud VMs. Risk simulation can be accelerated on the public cloud although the simulations do not scale well over the virtual cores of the instances; for example, for thirty two core instances no acceleration is achieved beyond sixteen cores. This results in wasted computational time per dollar spent on the simulation. Hence, maximum performance cannot be achieved thereby not fully exploiting the potential of the public cloud. Nevertheless, a 60x speedup is achieved on public instances over a baseline implementation. Interestingly, the private VMs are faster than the public instances. For example, the sequential CPU implementation, parallel CPU implementation and the parallel GPU implementation on private VMs are up to 40\%, 22\% and 16\% faster than the best performance achieved on public instances. 

\section{Conclusions}
\label{conclusion}
The cloud is a potential platform for performing ad hoc risk simulations important to financial applications. Scalable and on-demand resources of the cloud are attractive for running ad hoc simulations and for meeting their data, memory and computational requirements. The research reported in this paper was motivated towards experimentally verifying whether clouds are ready to accelerate financial simulations. A typical application employed in the financial industry, namely Aggregate Risk Analysis, was developed and deployed on a variety of cloud VMs. The implementation exploited parallelism to accelerate the application and efficient management of data and memory to accommodate the application on the cloud. The experimental results indicate that the application can be accommodated to run on the cloud and an acceleration of up to 60x over a baseline implementation can be obtained with hardware accelerators on the cloud. Nevertheless, there is poor efficiency in the acceleration achieved highlighting the inability to harness the full potential of all available compute cores resulting in wasted computational performance. It is noted that the private VMs perform better than the public VMs. 

Migrating financial applications onto the cloud is viable since the cloud provides a suitable platform to accommodate the computational, data and memory demands of ad hoc simulations. This is of significant benefit to the financial industry as well as its associated industries since the scalability and availability of resources on an on-demand basis reduce maintenance costs. However, while acceleration was achieved for the simulation, in our experience it could not be run most efficiently on the public cloud since there was wasted computational time for every dollar spent.

\end{document}